\documentclass[prd,aps,showpacs,notitlepage,superscriptaddress,nofootinbib]{revtex4-1}
\usepackage{epsfig}
\usepackage{color}
\usepackage{url}
\usepackage{amsmath}
\usepackage{multirow}

\def\bea#1\eea{\begin{align}#1\end{align}}
\def\lnb#1{{\rm ln} #1}

\newcommand{\bef}{\begin{figure}[htb]\centering}
\newcommand{\eef}{\end{figure}}
\newcommand{\rd}{{\rm d}}

\newcommand{\NLO}{\rm NLO}

\newcommand{\beq}{\begin{equation}}
\newcommand{\eeq}{\end{equation}}
\newcommand{\beqa}{\begin{eqnarray}}
\newcommand{\eeqa}{\end{eqnarray}}
\newcommand{\ds}{{\rm d}\hat{\sigma}}
\newcommand{\nn}{\nonumber}

\def\tauone{{\cal T}_1}

\def\tauonecut{{\cal T}_1^{cut}}

\begin{document}
\title{Inclusive jet production as a probe of polarized PDFs at a future EIC}


\author{Radja Boughezal}
\email{rboughezal@anl.gov}
\affiliation{High Energy Physics Division, Argonne National Laboratory, Argonne, Illinois 60439, USA}

\author{Frank Petriello}
\email{f-petriello@northwestern.edu}
\affiliation{High Energy Physics Division, Argonne National Laboratory, Argonne, Illinois 60439, USA}
\affiliation{Department of Physics and Astronomy, Northwestern University,  Evanston, Illinois 60208, USA}

\author{Hongxi Xing}
\email{hxing@northwestern.edu}
\affiliation{High Energy Physics Division, Argonne National Laboratory, Argonne, Illinois 60439, USA}
\affiliation{Department of Physics and Astronomy, Northwestern University,  Evanston, Illinois 60208, USA}

\begin{abstract}

We present a detailed phenomenological study of polarized inclusive jet production in electron-proton collisions at a future Electron-Ion Collider (EIC).  Our analysis is performed at next-to-leading order in perturbative QCD using the numerical code {\tt DISTRESS} and includes all relevant partonic channels and resolved photon contributions.  We elucidate the role of different kinematic regions in probing different aspects of proton and photon structure and study the impact of the EIC collision energy on the measurement of polarized parton distribution functions.  

\end{abstract}

\maketitle

\section{Introduction}

Among the most intriguing aspects of hadronic physics is the spin decomposition of the proton in terms of its partonic constituents. This has remained an outstanding puzzle for three decades~\cite{Ashman:1987hv,Ashman:1989ig}, and is one of the key motivations for the proposed Electron Ion Collider (EIC)~\cite{Accardi:2012qut}. This topic has attracted tremendous attention from both the experimental and theoretical QCD communities. To determine the contribution of quarks and gluons to the spin of the proton, according to the spin sum rule~\cite{Ji:1996ek, Jaffe:1989jz}, one needs to extract the helicity-dependent parton distribution functions inside the proton. A standard way to approach this goal is to perform a global QCD analysis of all available data taken in spin-dependent deep inelastic scattering (DIS) and proton-proton (pp) collisions, such as is done by the DSSV~\cite{deFlorian:2014yva} and NNPDF~\cite{Nocera:2014gqa} collaborations. The accuracy of these global fits relies upon the validity of QCD factorization and the high precision computation of the perturbative hard part coefficients. 

Collimated jets of hadrons are one of the main probes of the unpolarized partonic structure of the nucleon in current global fits~\cite{Harland-Lang:2014zoa,Dulat:2015mca,Ball:2017nwa}.  Due to the hard scale arising from the jet transverse momentum there are  negligible final state non-perturbative hadronic effect in jet measurements.  For the same reason, the double longitudinal spin asymmetry for jet production with large transverse momentum in DIS and pp collisions offer excellent sensitivity to the spin-dependent parton distribution functions of the individual quarks and gluons in the proton. In particular, a global analysis of the RHIC data~\cite{deFlorian:2014yva} based on next-to-leading order (NLO)  perturbative QCD on jet production in polarized proton-proton collisions has revealed clear evidence for a significant contribution of gluon spin to the proton spin in the large momentum fraction region. However, information from low momentum fractions is still missing due to the lack of data in this region. With the center-of-mass energies available at the proposed EIC, polarized electron-proton runs will open unique possibilities to study jet production for a wide range of inclusive jet transverse momentum and rapidity, which in principle can provide access to the low momentum fraction region. Considering as well the anticipated high luminosity of the EIC, we anticipate that it can shed light on the helicity-dependent parton distribution functions and provide a deeper understanding of the fundamental spin structure of the proton.  Similar analyses have shown that the EIC can play an important role in understanding the nuclear dependence of parton distribution functions (PDFs) ~\cite{Klasen:2017kwb,Klasen:2018gtb}.

Recently, a NLO computation of the double longitudinal spin asymmetry for inclusive jet production in polarized DIS has been performed in Ref.~\cite{Hinderer:2017ntk,Uebler:2017glm}, using the narrow cone approximation~\cite{Jager:2004jh} to enable a fully analytical calculation.  Previous work has studied aspects of inclusive jet photoproduction in polarized collisions~\cite{deFlorian:1999ge,Jager:2008qm}.  In this paper, we aim to provide a fully differential NLO Monte Carlo computation for this process, which allows any kinematic cut to be imposed on the final state. In particular, we implement the recently developed N-jettiness subtraction scheme~ \cite{Boughezal:2015dva, Gaunt:2015pea} extended to describe polarized collisions~\cite{Boughezal:2017tdd} in order to regularize all QCD infrared divergences.  We present a detailed phenomenological analysis of inclusive jet production in polarized electron-proton scattering at a future EIC.  We summarize below several key aspects and findings of our study.
\begin{itemize}

\item We study three possible collision energies for a future EIC to determine how they differ in their sensitivity to proton structure.  Higher collision energies generally offer sensitivity to more aspects of hadronic structure, particularly to the hadronic structure of the proton.

\item We include all partonic contributions relevant to inclusive jet production, including both direct and resolved photon contributions that become relevant when the final-state lepton travels collinear to the beam direction.  By including all relevant channels we are able to determine which regions of phase space are sensitive to different aspects of proton structure.

\item We perform a detailed study of the unpolarized cross section and double longitudinal spin asymmetry throughout the accessible kinematic range in tranverse momentum and pseudorapidity, and point out which regions are sensitive to which aspects of proton structure.  Particularly in higher-energy collisions, different regions of jet transverse momentum can be selected to probe either the resolved photon distributions or the helicity-dependent proton PDFs.

\item We estimate the effect of EIC statistical errors and current PDF errors on jet production.  The PDF errors are much larger than the estimated statistical ones over much of phase space, demonstrating the EIC potential to greatly improve our knowledge of PDFs.

\item We quantify the effect of changing the jet radius on EIC phenomenology, which turns out to be small. 

\end{itemize}
A primary finding of our study is that the observable $A_{LL}$ in inclusive jet production is very sensitive to the gluon helicity distribution, particularly in the high transverse momentum and forward rapidity region, thus providing a good channel to reduce the uncertainty in determine the gluon contribution to the proton spin.  Our analysis is complementary to other important studies that have demonstrated the sensitivity of EIC jet production to various other aspects of proton structure~\cite{Aschenauer:2015ata,Chu:2017mnm,Aschenauer:2017oxs}, as described later in the text. 

The rest of the paper is organized as follows.  We review our fixed-order perturbative QCD theoretical framework in Section~\ref{sec:framework} and detail all partonic channels included in our calculation.  All indicating scattering processes have been incoporated into the numerical code {\tt DISTRESS}~\cite{Abelof:2016pby}.  In Section~\ref{sec:numerics} we detail the numerical settings and parameter choices used in our study.  Numerical results for the three chosen EIC collision energies are presented in Sections~\ref{sec:141},~\ref{sec:63} and~\ref{sec:44}.  Finally, we summarize and conclude in Section~\ref{sec:conc}.

\section{Theoretical framework}
\label{sec:framework}

\begin{figure}[h]
\centering
\includegraphics[width=1.5in]{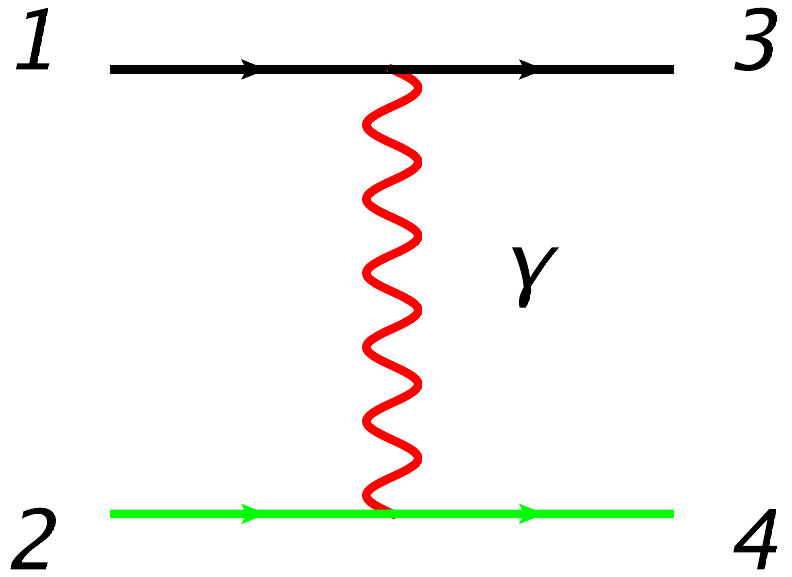}%
\caption{Feynman diagram for the leading-order process $q(p_1)+l(p_2) \to q(p_3)+l(p_4)$.  We have colored the photon line red, the lepton lines green and the quark lines black.} \label{fig:LOdiag}
\end{figure}

We sketch here the theoretical framework used in our study.  For more details we refer the reader to the discussion in Ref.~\cite{Abelof:2016pby}.   Our analysis is performed using fixed-order perturbative QCD through ${\cal O}(\alpha_s)$ in the strong coupling constant.  We include the leading contributions in the electromagnetic coupling, which go as ${\cal O}(\alpha^2)$.  We express the hadronic cross section in the following notation:
\begin{equation}
\rd\sigma = \rd\sigma_{\text{LO}}+\rd\sigma_{\NLO}+\ldots \,,
\end{equation}
where the ellipsis denotes neglected higher-order terms.  This generic equation holds for both the polarized and unpolarized cross sections.  The LO subscript refers to the ${\cal O}(\alpha^2)$ term while the NLO subscript denotes the ${\cal O}(\alpha^2\alpha_s)$ correction. For the partonic cross sections, we introduce superscripts that denote the powers of both $\alpha$ and $\alpha_s$ that appear.  For example, the leading quark-lepton scattering process is expanded as 
\begin{equation}
\rd\hat{\sigma}_{ql}= \rd\hat{\sigma}_{ql}^{(2,0)}+\rd\hat{\sigma}_{ql}^{(2,1)}+\ldots \,.
\end{equation}
Here, the $\rd\hat{\sigma}_{ql}^{(2,0)}$ denotes the ${\cal O}(\alpha^2)$ correction, while $\rd\hat{\sigma}_{ql}^{(2,1)}$ indicates the  ${\cal O}(\alpha^2\alpha_s)$ term.  The leading-order hadronic cross section can be written as a convolution of parton distribution functions with these partonic cross sections, 
\begin{equation}
\label{eq:sigLO}
\rd\sigma_{\text{LO}} = \int \frac{\rd \xi_1}{\xi_1} \frac{\rd \xi_2}{\xi_2} \sum_q \left[ f_{q/H}(\xi_1) f_{l/l}(\xi_2) \rd\hat{\sigma}_{ql}^{(2,0)} 
	+  f_{\bar{q}/H}(\xi_1) f_{l/l}(\xi_2) \rd\hat{\sigma}_{\bar{q}l}^{(2,0)}\right]. 
\end{equation} 
Here, $f_{q/H}(\xi_1)$ is the usual parton distribution function that describes the distribution of a quark $q$ in the hadron $H$ carrying a fraction $\xi_1$ of the hadron momentum.  $f_{l/l}(\xi_2)$ is the distribution for finding a lepton with momentum fraction $\xi_2$ inside the original lepton.  At leading order this is just $f_{l/l}(\xi_2)=\delta(1-\xi_2)$, but it is modified at higher orders in the electromagnetic coupling by photon emission.  The dependence of these distribution functions on the $\overline{\text{MS}}$ factorization scale $\mu_F$ is implicit. 
$d\hat{\sigma}_{ql}^{(2,0)}$ is the differential partonic cross section.  At leading order only the partonic channel $q(p_1)+l(p_2) \to q(p_3)+l(p_4)$ and the same process with anti-quarks contribute.  The relevant Feynman diagram is shown in Fig.~\ref{fig:LOdiag}.

\begin{figure}[h]
\centering
\includegraphics[width=4.5in]{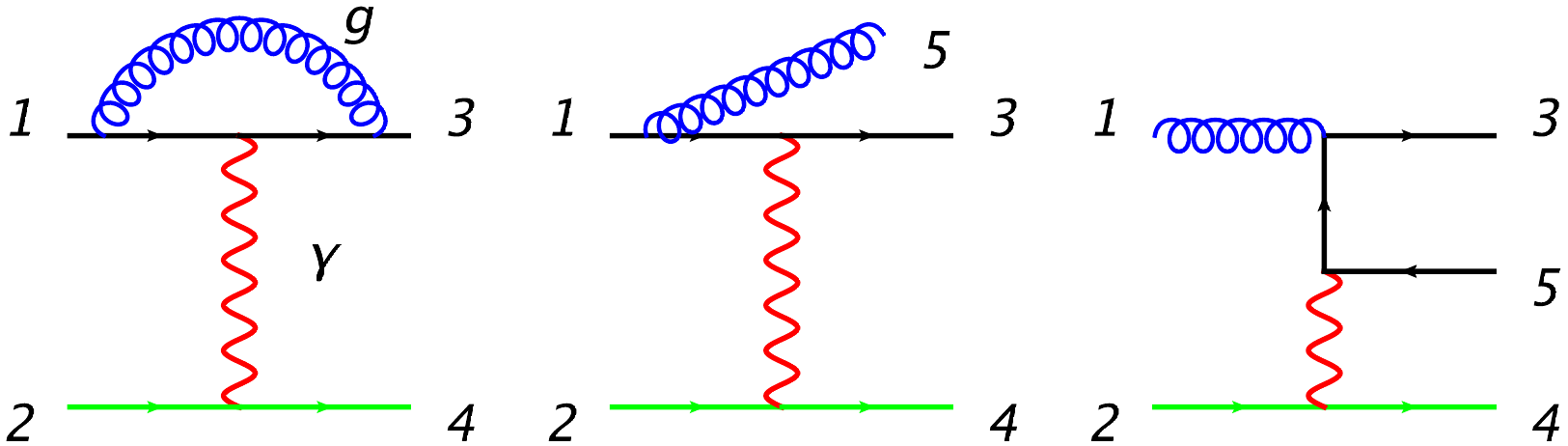}%
\caption{Representative Feynman diagrams contributing to the perturbative QCD corrections at NLO: virtual corrections to the 
$q(p_1)+l(p_2) \to q(p_3)+l(p_4)$ process (left); real emission correction $q(p_1)+l(p_2) \to q(p_3)+l(p_4)+g(p_5)$ (middle); the process $g(p_1)+l(p_2) \to q(p_3) +l(p_4)+\bar{q}(p_5)$ (right).  We have colored the photon line red, the lepton lines green, the gluon lines blue and the quark lines black.} \label{fig:NLOlepdiag}
\end{figure}

At the next-to-leading order level several new structures appear.  The leading-order  quark-lepton scattering channel receives both virtual and real-emission corrections that are separately infrared divergent.  We have performed two calculations using dipole subtraction~\cite{Catani:1996vz} and $N$-jettiness subtraction~\cite{Boughezal:2015dva,Gaunt:2015pea} to regularize and cancel these divergences.  The agreement we find between these two approaches serves as a check of our result.  Initial-state collinear divergences are absorbed into PDFs via mass factorization.   At this order in perturbation theory a gluon-lepton scattering channel also contributes.  The collinear divergences that appear in this channel are removed by mass factorization.  Representative Feynman diagrams for these processes are shown in Fig.~\ref{fig:NLOlepdiag}.

\begin{figure}[h]
\centering
\includegraphics[width=3.0in]{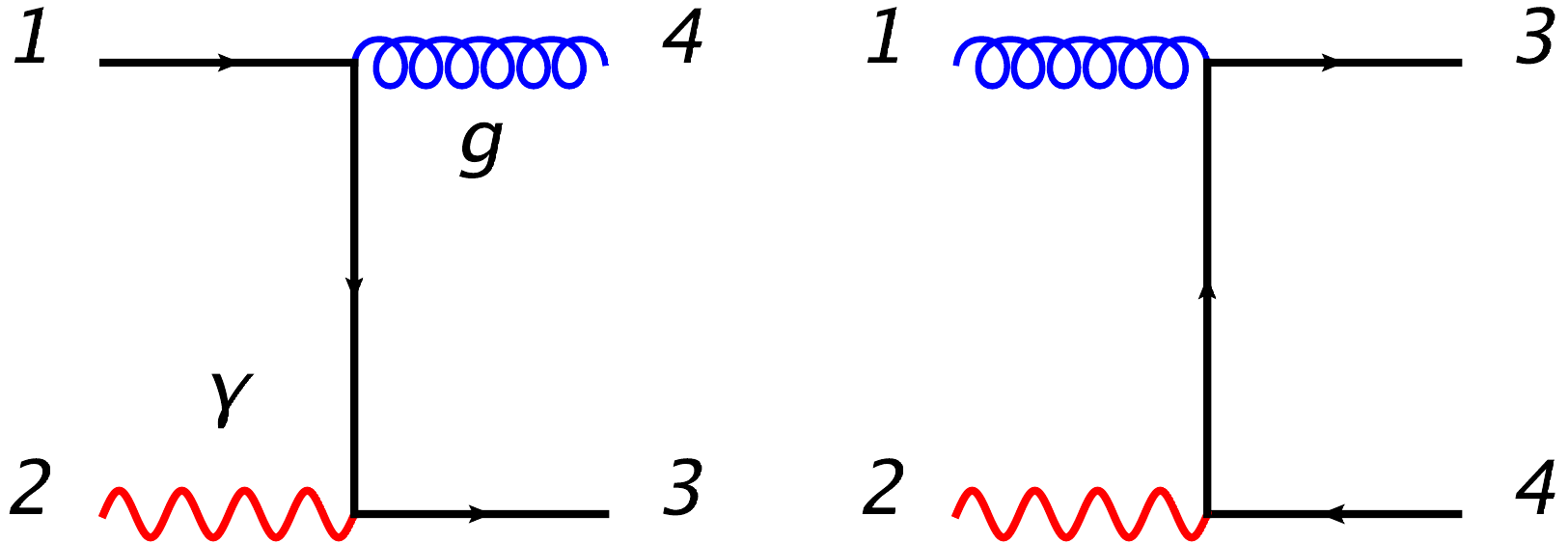}%
\caption{Representative Feynman diagrams contributing to the $q(p_1)+\gamma(p_2) \to q(p_3)+g(p_4)$ (left) and $g(p_1)+\gamma(p_2) \to q(p_3)+\bar{q}(p_4)$ scattering processes.} \label{fig:NLOphotdiag}
\end{figure}

We can express the NLO hadronic cross section in the following form:
\begin{equation}
\rd\sigma_{\NLO}=\int\frac{\rd\xi_1}{\xi_1}\frac{\rd\xi_2}{\xi_2}\bigg\{f_{g/H}^1f_{l/l}^2 \ds_{gl}^{(2,1)} + f_{g/H}^1f_{\gamma/l}^2\ds_{g\gamma}^{(1,1)} 
 +\sum_{i=q,\bar{q}}\bigg[f_{i/H}^1f_{l/l}^2\ds_{il}^{(2,1)}+f_{i/H}^1f_{\gamma/l}^2\ds_{i\gamma}^{(1,1)}\bigg]\bigg\},
\end{equation}
where we have abbreviated $f_{i/j}^k=f_{i/j}(\xi_k)$.  The contributions $\ds_{gl}^{(2,1)}$, $\ds_{ql}^{(2,1)}$ and $\ds_{\bar{q}l}^{(2,1)}$ denote the usual deep-inelastic scattering (DIS) partonic channels computed to NLO in QCD with zero lepton mass.  The terms $\ds_{q\gamma}^{(1,1)}$, $\ds_{\bar{q}\gamma}^{(2,1)}$ and $\ds_{g\gamma}^{(1,1)}$ denote new contributions arising when $Q^2 = -(p_2-p_4)^2 \approx 0$. These are associated with a virtual photon that is nearly on-shell, and a final-state lepton that travels down the beam pipe. The transverse momentum of the leading jet is balanced by the additional jet present in these diagrams, and the final-state lepton is not observed.  This kinematic configuration leads to a QED collinear divergence for vanishing lepton mass.  While it is physically regulated by the lepton mass, it is more convenient to obtain these corrections by introducing a photon distribution function in analogy with the usual parton distribution function.  The collinear divergences that appear in the matrix elements computed with vanishing lepton mass can be absorbed into this distribution function.  This procedure is described for inclusive jet production in Refs.~\cite{Hinderer:2015hra,Abelof:2016pby}.  Representative diagrams for the photon-initiated processes are shown in Fig.~\ref{fig:NLOphotdiag}.

Since it is our primary calcutional tool we give here a brief description of the $N$-jettiness subtraction technique as introduced in Ref.~\cite{Boughezal:2015dva}.  The starting point of this method is the $N$-jettiness event shape variable~\cite{Stewart:2010tn}, defined in the one-jet case of current interest as
\begin{equation}
\label{eq:taudef}
\tauone = \frac{2}{Q^2} \sum_i \text{min}\left\{p_B \cdot q_i, p_J \cdot q_i \right\}.
\end{equation}
Here, $p_B$ and $p_J$ are light-like four-vectors along the initial-state hadronic beam and final-state jet directions, respectively.  The $q_i$ denote the four-momenta of all final-state partons.   Values of $\tauone$ near zero indicate a final state containing a single narrow energy deposition, while larger values denote a final state containing two or more well-separated energy depositions.  Restricting $\tauone >0$ removes all singular limits of the quark-lepton matrix elements, for example when the additional parton that appear in the real emission corrections is soft or collinear to the beam or the final-state jet.  This can be seen from Eq.~(\ref{eq:taudef}); if $\tauone >0$ then $q_i$ must be resolved.  Since all unresolved limits are removed, the ${\cal O}(\alpha^2\alpha_s)$ correction in this phase space region can be obtained from a leading-order calculation of two-jet production in electron-nucleon collisions.  When $\tauone$ is smaller than any other hard scale in the problem, it can be resummed to all orders in perturbation theory~\cite{Stewart:2009yx,Stewart:2010pd}.  Expansion of this resummation formula to ${\cal O}(\alpha^2\alpha_s)$ gives the NLO correction to the quark-lepton scattering channel for small $\tauone$.

To obtain the full NLO result using this idea we partition the phase space 
into regions above and below a cutoff on $\tauone$, which we label $\tauonecut$:
\begin{equation}
\label{eq:partition}
\begin{split}
\rd\sigma_{ql}^{(2,1)} &= \int {\rm d}\Phi_{\text{V}} \, |{\cal M}_{\text{V}}|^2 
+\int {\rm d}\Phi_{\text{R}} \, |{\cal M}_{\text{R}}|^2 \, \theta_1^{<}
+\int {\rm d}\Phi_{\text{R}} \, |{\cal M}_{\text{R}}|^2 \, \theta_1^{>} \\
 \equiv & \rd\sigma_{ql}^{(2,1)}(\tauone < \tauonecut)+\rd\sigma_{ql}^{(2,1)}(\tauone > \tauonecut)
\end{split}
\end{equation}
We have abbreviated $\theta_1^{<} = \theta(\tauonecut-\tauone)$ and $\theta_1^{>} = \theta(\tauone-\tauonecut)$ and have used the notation R and V to denote real and virtual corrections to the cross section.  The first two terms in this expression all have $\tauone<\tauonecut$, and have been collectively denoted as $\rd\sigma_{ql}^{(2,1)}(\tauone < \tauonecut)$.  The remaining term has $\tauone>\tauonecut$, and have been collectively denoted as $\rd\sigma_{ql}^{(2,1)}(\tauone > \tauonecut)$.   We obtain $\rd\sigma_{ql}^{(2,1)}(\tauone > \tauonecut)$ from a LO calculation of two-jet production.  This is possible since no unresolved limit occurs in this phase-space region.  We derive  $\rd\sigma_{ql}^{(2,1)}(\tauone < \tauonecut)$ using the all-orders resummation of this process~\cite{Stewart:2009yx,Stewart:2010pd}.

\begin{figure}[h]
\centering
\includegraphics[width=4.5in]{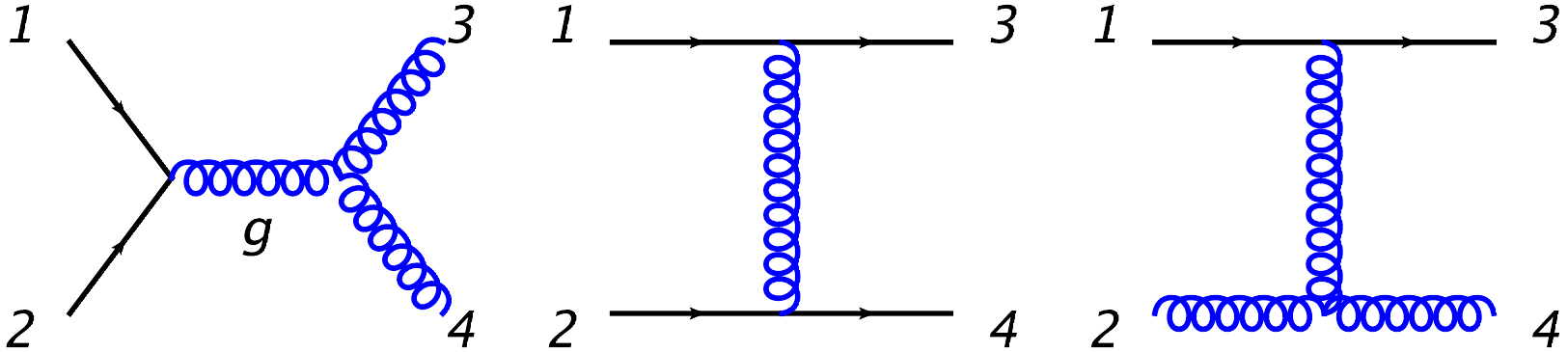}%
\caption{Representative Feynman diagrams contributing to the $q(p_1)+\bar{q}(p_2) \to g(p_3)+g(p_4)$ (left), $q(p_1)+q^{\prime}(p_2) \to q(p_3)+q^{\prime}(p_4)$ (middle), and $q(p_1)+g(p_2) \to q(p_3)+g(p_4)$ (right) scattering processes.} \label{fig:NNLO2jet}
\end{figure}

Finally, nearly on-shell photons can also lead to resolved photon contributions to the cross section, in which these photons split into QCD partons that enter the hard-scattering process.  Since this splitting occurs for low virtualities this process receives important non-perturbative QCD contributions.  Although the hard scattering corrections are formally ${\cal O}(\alpha_s^2)$, they can be sizeable due to the non-perturbative distribution of the partons inside the photon.   We write these distributions as $f_{a/\gamma}(x,\mu)$ and $\Delta f_{a/\gamma}(x,\mu)$ in the unpolarized and polarized cases respectively.  This leads to corresponding non-perturbative parton-in-lepton distributions.  In the unpolarized case we have
\begin{equation}
\label{eq:partinlep}
f_{a/l}(x,\mu) = \int_x^1 \frac{dy}{y} P_{\gamma l}(y,\mu) f_{a/\gamma}\left(\frac{x}{y},\mu\right).
\end{equation}
The function $ P_{\gamma l}$ in the leading-logarithmic approximation is given by 
\begin{equation}
 P_{\gamma l}(y,\mu) = \frac{\alpha}{2\pi}\frac{1+(1-y)^2}{y}\left[\lnb{\frac{\mu^2}{y^2 m_l^2}}-1\right].
\end{equation}
In the polarized case these formulae take the form
\begin{equation}
\Delta f_{a/l}(x,\mu) = \int_x^1 \frac{dy}{y} \Delta P_{\gamma l}(y,\mu) \Delta f_{a/\gamma}\left(\frac{x}{y},\mu\right), \;\;\;
\Delta P_{\gamma l}(y,\mu) =\frac{\alpha}{2\pi} \frac{2-y}{y}\left[\lnb{\frac{\mu^2}{y^2 m_l^2}}\right].
\end{equation}
We will discuss the form of the non-perturbative parton distributions of the photon later in this manuscript.  We can write the resolved-photon contribution to the unpolarized cross section as
\begin{eqnarray}
\label{eq:hxsecnnlo}
\rd\sigma_{{\rm res}}&=&\int\frac{\rd\xi_1 \rd\xi_2}{\xi_1 \xi_2}\bigg\{
\sum_{i=q,\bar{q}}\bigg[f_{g/H}^1f_{i/l}^2\,\ds_{gi}^{(0,2)}
+f_{i/H}^1f_{g/l}^2\,\ds_{ig}^{(0,2)}\bigg] + f_{g/H}^1f_{g/l}^2\,\ds_{gg}^{(0,2)} \nn\\
&+& \sum_q\bigg[f_{q/H}^1f_{\bar{q}/l}^2\,\ds_{q\bar{q}}^{(0,2)}+f_{\bar{q}/H}^1f_{q/l}^2\,\ds_{\bar{q}q}^{(0,2)} 
+f_{q/H}^1f_{q/l}^2\,\ds_{qq}^{(0,2)}+f_{\bar{q}/H}^1f_{\bar{q}/l}^2\,\ds_{\bar{q}\bar{q}}^{(0,2)} 
\bigg] \phantom{\sum_q} \nn \\
&+&\sum_{q \ne q'}\bigg[f_{q/H}^1f_{q'/l}^2\,\ds_{qq'}^{(0,2)}+f_{\bar{q}/H}^1f_{\bar{q}'/l}^2\,\ds_{\bar{q}\bar{q}'}^{(0,2)}  
+f_{q/H}^1f_{\bar{q}'/l}^2\,\ds_{q\bar{q}'}^{(0,2)}+f_{\bar{q}/H}^1f_{q'/l}^2\,\ds_{\bar{q}q'}^{(0,2)}\bigg]\bigg\},
\end{eqnarray}
where we have implicitly set the arguments of $f^1_{i/H} (\xi_1,\mu) = f^1_{i/H}$ and $f^2_{i/l} (\xi_2,\mu) = f^2_{i/l}$.  A similar expression holds in the polarized case.  Although the partonic scattering cross sections go like ${\cal O}(\alpha_s^2)$ as the superscripts in Eq.~(\ref{eq:hxsecnnlo}) indicate, it can be shown that in certain limits the $f_{i/\gamma}$ distributions go like $1/\alpha_s$~\cite{Klasen:2002xb}.  We consequently count this contribution as part of the next-to-leading order result. Representative diagrams contributing to the partonic scattering processes of Eq.~(\ref{eq:hxsecnnlo}) are shown in Fig.~\ref{fig:NNLO2jet}.  The total cross section becomes
\begin{equation}
\rd\sigma_{\text{tot}} = \rd\sigma_{\text{LO}}+\rd\sigma_{\NLO}+ \rd\sigma_{{\rm res}}.
\end{equation}

All of these partonic channels have been incorporated into the numerical program {\tt DISTRESS}~\cite{Abelof:2016pby}, which predicts both polarized and unpolarized cross sections for jet production in DIS.  A summary of the partonic contributions in {\tt DISTRESS} used in this analysis, the kinematic range in which they contribute, and the PDFs to which they are sensitive are shown in Table~\ref{tab:chan}.

\begin{table}[h!]
\centering
\begin{tabular}{|c||c|c|}
\hline
Partonic channel & $Q^2$ region & Contributing PDFs \\
\hline
\hline
$ql$ & $Q^2>0$ & $f_{q/H},\Delta f_{q/H}$ \\
$gl$ & $Q^2>0$ & $f_{g/H},\Delta f_{g/H}$ \\
$q\gamma$ & $Q^2 \approx 0$ & $f_{q/H},f_{\gamma /l},\Delta f_{q/H},\Delta f_{\gamma /l}$ \\
$g\gamma$ & $Q^2 \approx 0$ & $f_{g/H},f_{\gamma /l},\Delta f_{g/H},\Delta f_{\gamma /l}$ \\
$qq$ & $Q^2 \approx 0$ & $f_{q/H},f_{q/ \gamma},\Delta f_{q/H},\Delta f_{q/\gamma}$ \\
$qg$ & $Q^2 \approx 0$ & $f_{q/H},f_{q/ \gamma},\Delta f_{q/H},\Delta f_{q/\gamma},f_{g/H},f_{g/ \gamma},\Delta f_{g/H},\Delta f_{g/\gamma}$ \\
$gg$ & $Q^2 \approx 0$ & $f_{g/H},f_{g/ \gamma},\Delta f_{g/H},\Delta f_{g/\gamma}$ \\
\hline
\end{tabular}
\caption{Summary of partonic channels, the region of photon virtuality $Q^2$ for which they contribute, and the distribution functions to which they are sensitive.}
\label{tab:chan}
\end{table}

\section{Numerical setup}
\label{sec:numerics}

We describe in this section the numerical parameters and settings used in our study.  For convenience we summarize below in Table~\ref{tab:kin} the various kinematic quantities used in our analysis.  We reconstruct jets using the anti-$k_T$ algorithm~\cite{Cacciari:2008gp}.  Unless noted otherwise we use a jet radius $R=0.8$.  The transverse momenta and pseudorapidities of the jets are reconstructed in the center-of-mass frame of the electron-proton scattering process.  We set the renormalization and factorization scales to the transverse momentum of the jet, $\mu_R=\mu_F=p_T^{j}$.  A detailed study of the theoretical uncertainty arising from scale variation of the cross section was performed previously~\cite{Abelof:2016pby}, with the conclusion that once the NNLO corrections are incorporated the scale dependence is reduced to the few-percent level.  Since we do not expect this to be a limiting uncertainty by the time of EIC data-taking we do not consider the scale dependence further here.

\begin{table}[h!]
\centering
\begin{tabular}{|c|c|}
\hline
\hline
$\sqrt{s}$ & Center-of-mass energy of the proton-lepton collision\\
$ Q^2 = -(p_2 - p_4)^2$ & Virtuality of the photon exchanged in the DIS process\\
$ p_T^j$ & Tranverse momentum of the observed jet in the lab frame\\
$ \eta^j$ & Pseudorapidity of the observed jet in the lab frame\\
\hline
\end{tabular}
\caption{Definition of kinematic parameters used in our study.}
\label{tab:kin}
\end{table}

For the unpolarized parton distributions in the proton we use the NNPDF 3.1 PDFs~\cite{Ball:2017nwa} extracted at next-to-leading order in QCD perturbation theory.  To describe the polarized parton content of the proton we use the NNPDF 1.1 polarized PDFs~\cite{Nocera:2014gqa} unless noted otherwise.  The one-sigma PDF uncertainties shown in the following sections are obtained by evaluating the cross section for the 100 replica sets provided by NNPDF and combining their differences from the reference set according to the standard methodology~\cite{Ball:2008by}.
For the non-perturbative unpolarized parton distributions of the photon we use the leading-order GRV distributions from Ref.~\cite{Gluck:1991jc}.  The corresponding polarized distributions have not been determined from data and require modeling, as discussed in Ref.~\cite{Gluck:1992fy}.  We study both the minimal and maximal models from this reference, which correpsond to different choices for the boundary conditions used when solving the evolution equations which these distributions satisfy.\footnote{We thank W.Vogelsang for providing numerical routines for the polarized photon distributions.}  We note that all considered PDFs are defined in the $\overline{\text{MS}}$ factorization scheme.

In order to estimate the sensitivity of inclusive jet production to the polarized structure of the proton for different EIC realizations, we consider three different setups corresponding to different center-of-mass scattering energies~\cite{Aschenauer:2017jsk}.  These different energies, together with the associated ranges of jet transverse momenta and pseudorapidities considered, are shown in Table~\ref{tab:eic}.  We assume 10 fb$^{-1}$ of integrated luminosity for all design parameters.  For simplicity we also assume 100\% polarization for both the initial-state electrons and protons.  The results we obtain can be simply rescaled to account for the polarization fractions eventually realized. To illustrate graphically what inclusive jet production at an EIC teaches us about proton structure, we show in Fig.~\ref{fig:xrange} how the different measured $(p_T^j, \eta^j)$ regions map into the Bjorken-$x$ and $Q^2$ ranges of the PDFs.  We assume leading-order $2 \to 2$ kinematics in order to make these plots.  We see that particularly at high center-of-mass energies that low Bjorken-$x$ can be probed. 

\begin{figure}[h]
\psfig{file=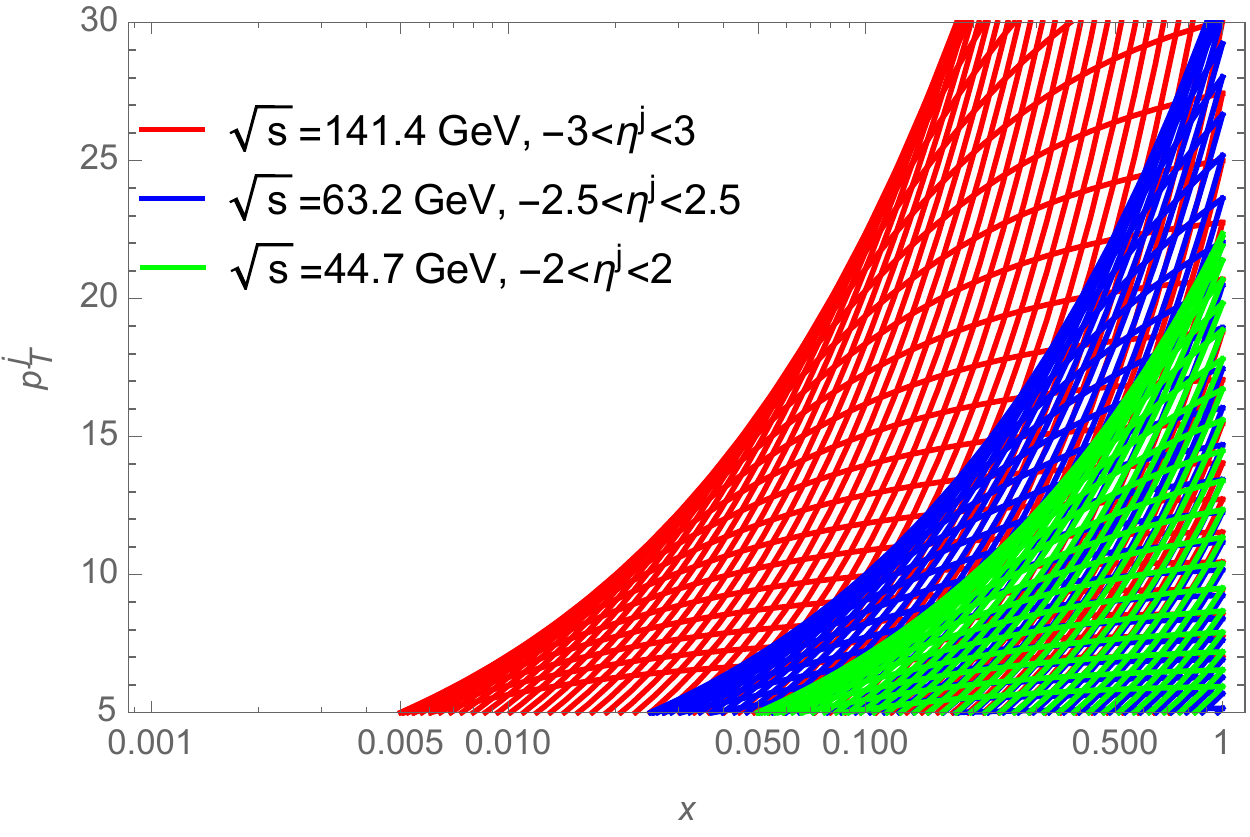, width=4.5in}
\caption{Ranges of Bjorken-$x$ and $Q^2$ probed by inclusive jet measurements for each studied scattering energy.}
\label{fig:xrange}
\eef

We note that the experimental cuts considered allow for $Q^2 \approx 0$, where the final-state lepton goes down the beam pipe and is not observed.  This kinematic configuration allows for on-shell photons and is responsible for the parton-photon and parton-parton scattering channels that appear in the previous section.  In order to investigate the effects of these channels we will also consider the effect of a $Q^2> 10 \, {\rm GeV}^2$ cuts for the energy $\sqrt{s}=141.4 \, {\rm GeV}$.  Such a cut removes the on-shell photon contributions. 

\begin{table}[h!]
\centering
\begin{tabular}{|c|c|c|}
\hline
 $\sqrt{s}$ & $p_T^{jet}$ range & $\eta^jet$ range \\
\hline
\hline
 141.4 GeV & $5 \, {\rm GeV} \leq p_T^{j} \leq 35 \, {\rm GeV}$ & $-3 \leq \eta^{j} \leq 3$ \\ 
 63.2 GeV & $5 \, {\rm GeV} \leq p_T^{j} \leq 30 \, {\rm GeV}$ & $-2.5 \leq \eta^{j} \leq 2.5$ \\ 
 44.7 GeV &  $5 \, {\rm GeV} \leq p_T^{j} \leq 20 \, {\rm GeV}$ & $-2 \leq \eta^{j} \leq 2$ \\ 
\hline
\end{tabular}
\label{tab:eic}
\caption{Considered run scenarios for a future EIC including center-of-mass scattering energies, jet transverse momentum ranges and jet pseudorapidity ranges.}
\end{table}

\section{Results for $\sqrt{s} = 141.4$ GeV}
\label{sec:141}

We begin by presenting numerical results for the largest center-of-mass energy considered in our study, $\sqrt{s} = 141.4$ GeV.  For this setting and for all other numerical results we consider both the total unpolarized cross section and the double-longitudinal spin asymmetry defined in Eq.~(\ref{eq:polasym}).  We study both quantities as functions of the jet transverse momentum and pseudorapidity.  Both inclusive jet production without and with a tagged lepton are considered for this collider energy.

\subsection{Inclusive jet production without a tagged lepton}

We begin by presenting the unpolarized cross section as a function of both the jet transverse momentum and pseudorapidity in Fig.~\ref{fig:sqrts141-totcr}.  No cut on the momentum transfer $Q^2$ is imposed, so that nearly on-shell photons contribute to the measured jet distributions.  The red bands in these plots show the PDF uncertainties as computed using the NNPDF 3.1 error sets.  Also shown are the estimated statistical errors at a future EIC assuming 10 fb$^{-1}$ of integrated luminosity.  The contributions to the cross section coming from resolved photons are shown separately in these plots as dashed lines.  In the lower panels the results are normalized to the central value of the predictions in order to more clearly illustrate the errors.  We see that the estimated PDF uncertainties are quite small, at or below the 1\% level over most of the accessible kinematics.  This is not surprising, as the unpolarized PDFs have been very well determined from a combination of HERA, LHC and lower energy data.  The estimated experimental statistical errors assuming 10 fb$^{-1}$ of integrated luminosity are also at or below the 1\% level except at high $p_T^{j}$ or in the high pseudorapidity regions.  We note that we have not attempted to estimate the experimental systematic errors in our study.  The resolved photon contribution to the cross section is important at low transverse momentum.  It falls off rapidly as $p_T^j$ is increased, suggesting that determinations of this quantity should focus on the low transverse momentum region to enhance its importance as compared to other partonic contributions.  

\begin{figure}[h]
\psfig{file=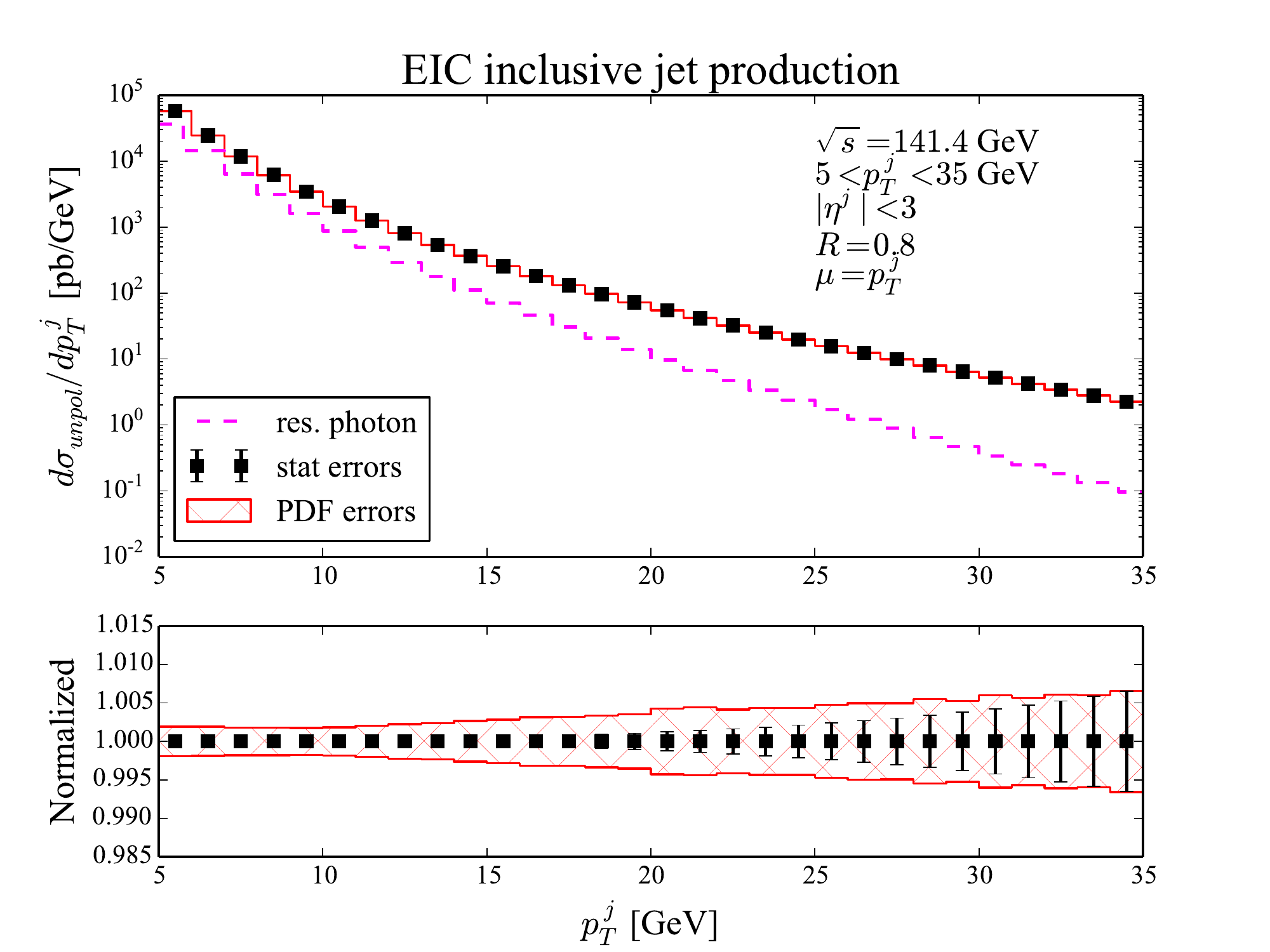, width=3.3in}
\hskip 0.2in
\psfig{file=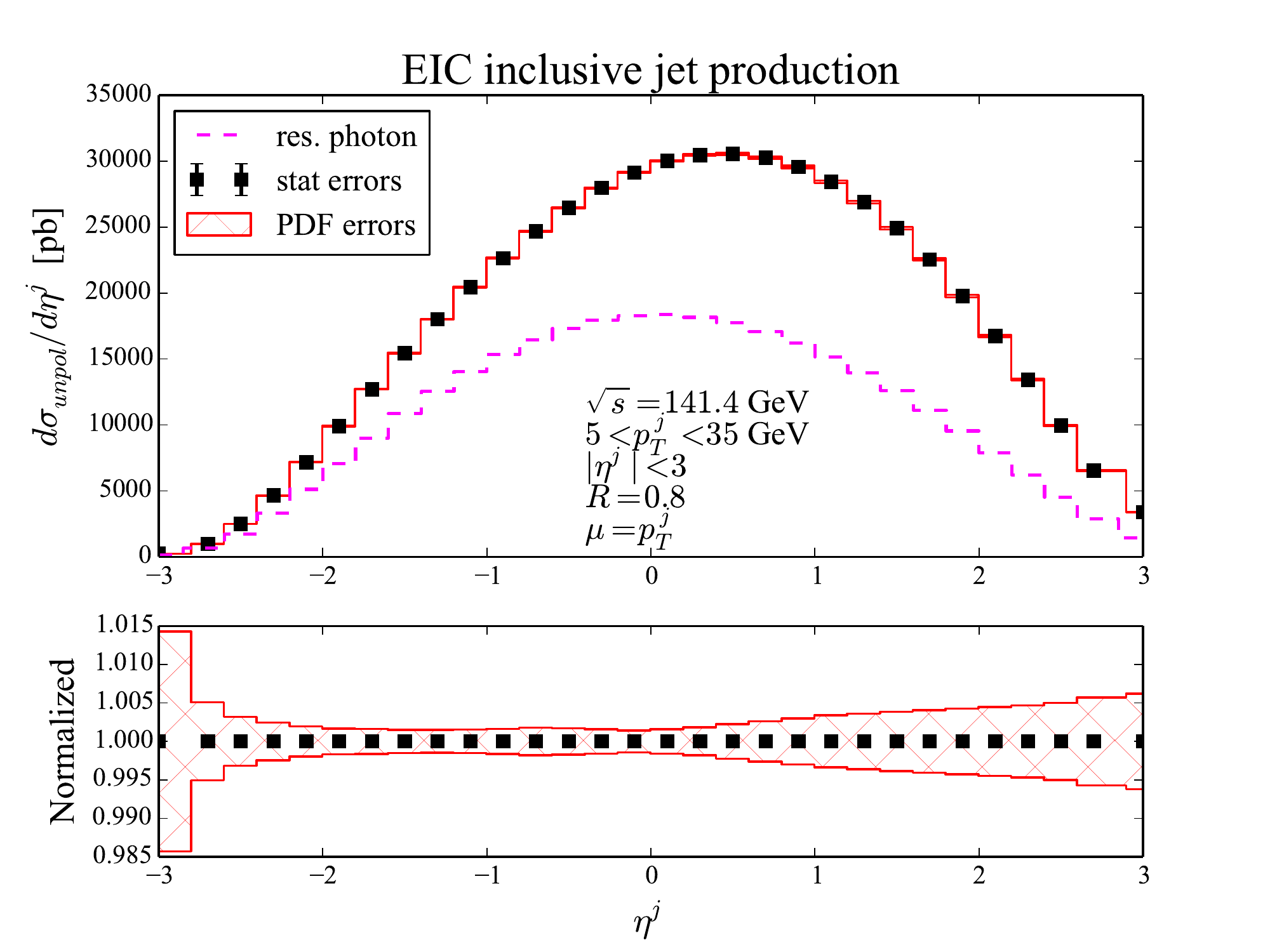, width=3.3in}
\caption{Total unpolarized cross section as a function of jet transverse momentum (left panel) and jet pseudorapidity (right panel). The resolved photon contribution is shown separately in the upper panel of each plot.  The lower panels normalize the results to the central values in order to more clearly illustrate the errors.}
\label{fig:sqrts141-totcr}
\eef

To make the partonic structure of these results more clear we show in Fig.~\ref{fig:sqrts141-totcr-split} the central values for each distribution split into the separate partonic channels.  The labeling of the partonic channels follows that introduced in Section~\ref{sec:framework}: the $ql$ and $gl$ channels denote diagrammatic contributions of the form shown in Figs.~\ref{fig:LOdiag} and~\ref{fig:NLOlepdiag} where the virtual photon exchanged between the quark and lepton line is off-shell, $q\gamma$ and $g\gamma$ denote contributions of the sort shown in Fig.~\ref{fig:NLOphotdiag} in which the photon is nearly on-shell and directly interacts with a parton coming from the proton, and the resolved contribution denotes terms where the photon splits at low virtuality into a parton before entering the hard scattering, as shown in Fig.~\ref{fig:NNLO2jet}.  We first discuss the structure of the transverse momentum distribution.  The resolved photon contributions dominate at low $p_T^j$ and fall off rapidly as $p_T^j$ is increased.  This occurs because of the multiple collinear splittings needed to obtain the parton from the initial lepton as shown in Eq.~(\ref{eq:partinlep}), leading to softer distributions for $f_{q/l}$ and $f_{g/l}$.  At intermediate values $p_T^j \sim 15-20$ GeV both the $ql$ and resolved channels are important, while at high $p_T^j$ the $ql$ channel dominates.  The $q\gamma$ and $g\gamma$ channels are smaller than the leading channel for all $p_T^j$.  The $gl$ channel is negligible throughout phase space. The $\eta^j$ distribution is dominated throughout phase space by the resolved photon channel.  This is not surprising as the total event rate is dominated by low-$p_T^j$ where this channel is largest.  We note that the $g\gamma$ channel becomes important at high $\eta^j$.  The sensitivity of the unpolarized jet production cross section to the resolved photon structure has been studied previously~\cite{Chu:2017mnm}, where the possibility of flavor-tagging to resolve the quark and gluon distributions has also been discussed.

\begin{figure}[h]
\psfig{file=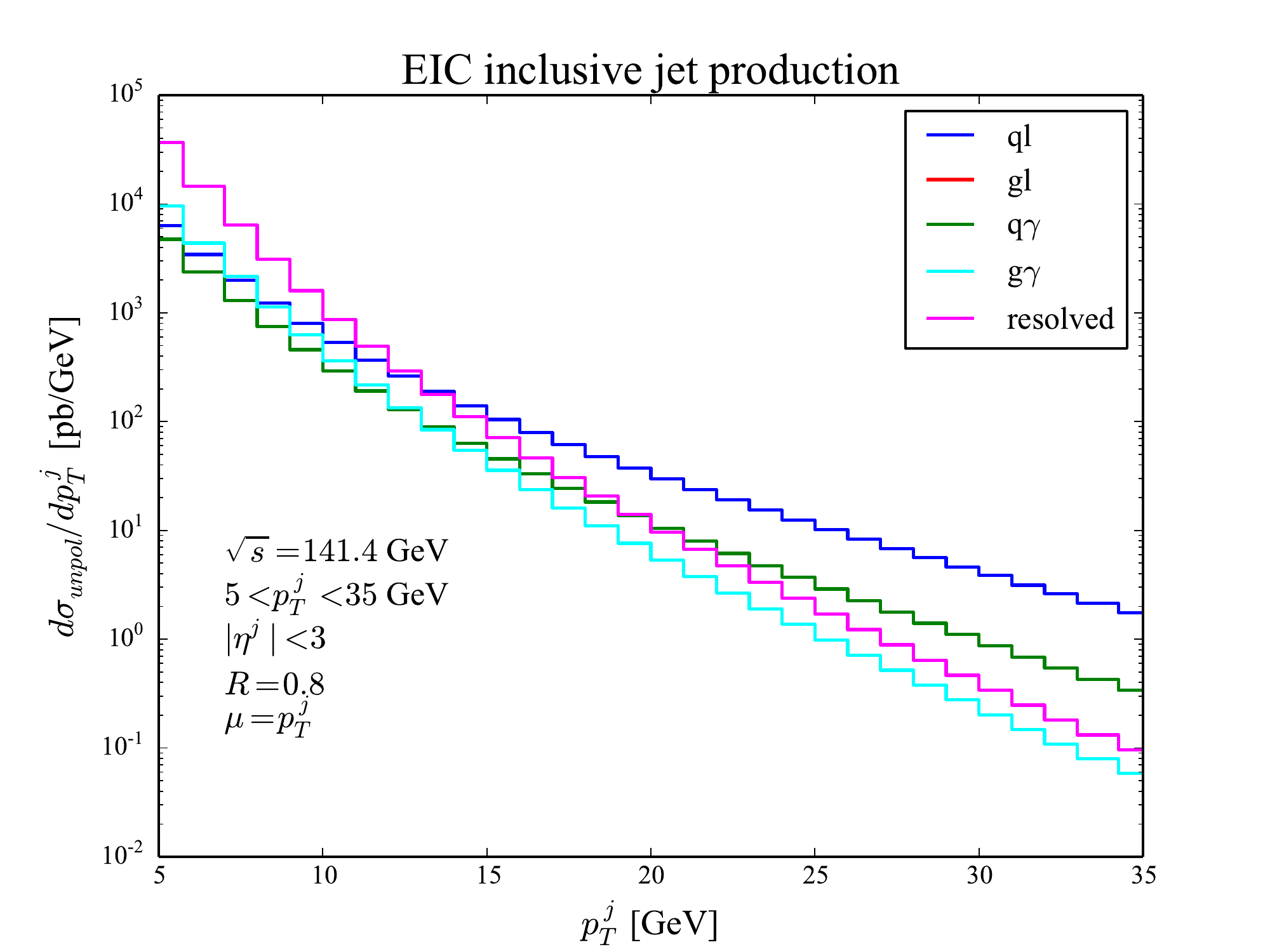, width=3.3in}
\hskip 0.2in
\psfig{file=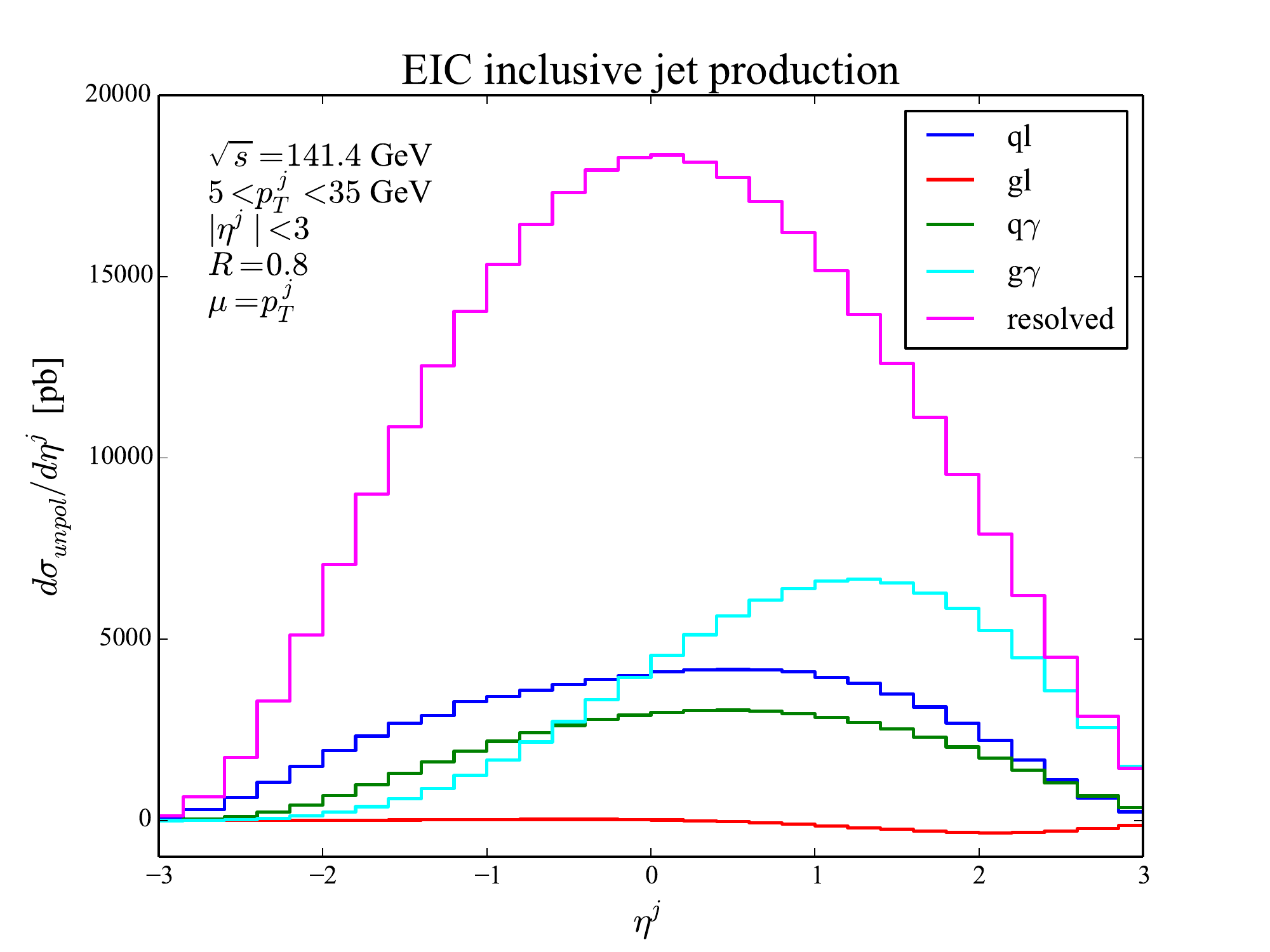, width=3.3in}
\caption{Split of the unpolarized transverse momentum and pseudorapidity distributions into partonic channels as described in the text.}
\label{fig:sqrts141-totcr-split}
\eef

We now study the double-longitudinal spin asymmetry, defined as
\begin{equation}
{\cal A}_{LL}=\frac{ d \sigma^{++}-d\sigma^{+-}-d\sigma^{-+}+d\sigma^{--}}{d\sigma^{++}+d\sigma^{+-}+d\sigma^{-+}+d\sigma^{--}}\, ,
\label{eq:polasym}
\end{equation}
where the first superscript refers to the helicity of the lepton and the second one to the proton.  The possibility of polarized beams at an EIC makes the measurement of this observable possible, allowing access to the polarized structure of the proton.  The spin asymmetry as a function of jet transverse momentum and jet pseudorapidity is shown in Fig.~\ref{fig:sqrts141-ALL}.  We again show the PDF errors and statistical errors for each distribution.  The PDF errors are significantly larger than in the unpolarized case, indicating the poorer understanding of the polarized structure of the proton.  The asymmetry increases as a function of $p_T^j$, reaching nearly 20\% for $p_T^j=35 $ GeV.  It is small throughout the studied $\eta^j$ range, since the event rate when integrated over $p_T^j$ is dominated by low transverse momentum where $A_{LL}$ is small.  The fact that the PDF errors are much larger than the estimated statistical errors over all of phase space shows that the EIC has the potential to greatly improve our knowledge of these distributions.  The increase of the PDF errors at low transverse momentum is due to the large uncertainty in the polarized PDFs at low Bjorken-$x$.

We show three additional quantities in the upper panels of each plot.  First, we show the resolved photon contribution to the asymmetry for both the minimal and maximal models of the polarized distribution functions of the photon defined in Ref.~\cite{Gluck:1992fy}.  We define these contributions by keeping only the resolved photon terms in the numerator of Eq.~(\ref{eq:polasym}), while keeping all contributions to the denominator.  The resulting quantity is directly proportional to the polarized photon distribution function.  Both models of the resolved photon distribution give small contributions to the asymmetry except at low values of $p_T^j$, indicating that intermediate and high-$p_T^j$ jet production is not sensitive to this distribution.  The situation is different for the $\eta^j$ distribution.  The maximal model gives nearly all of the asymmetry in the negative $\eta^j$ region, while the minimal model is small throughout the entire $\eta^j$ range.  This indicates that the uncertainties arising from the polarized photon distributions are large in the $\eta^j$ distribution.

We also show in these plots the results where the polarized gluon distribution is set to zero.  These are obtained by setting all numerator terms containing the gluon distribution to zero in Eq.~(\ref{eq:polasym}), while keeping all contributions in the denominator.  As the determination of the polarized gluon is a major goal of the EIC it is interesting to study the sensitivity of jet production to this important quantity.  The result obtained without the polarized gluon contribution differs by more than the estimated errors throughout the region $15 \, \text{GeV} \leq p_T^j \leq 35 \, \text{GeV}$.  The shape of $A_{LL}$ as a function of $\eta^j$ is qualitatively different when the polarized gluon is turned off.  Although the estimated statistical error indicates that this effect may be observable, the smallness of the asymmetry and the uncertainties in the resolved photon distribution indicates that such a determination of the polarized gluon may be difficult.  

\begin{figure}[h]
\psfig{file=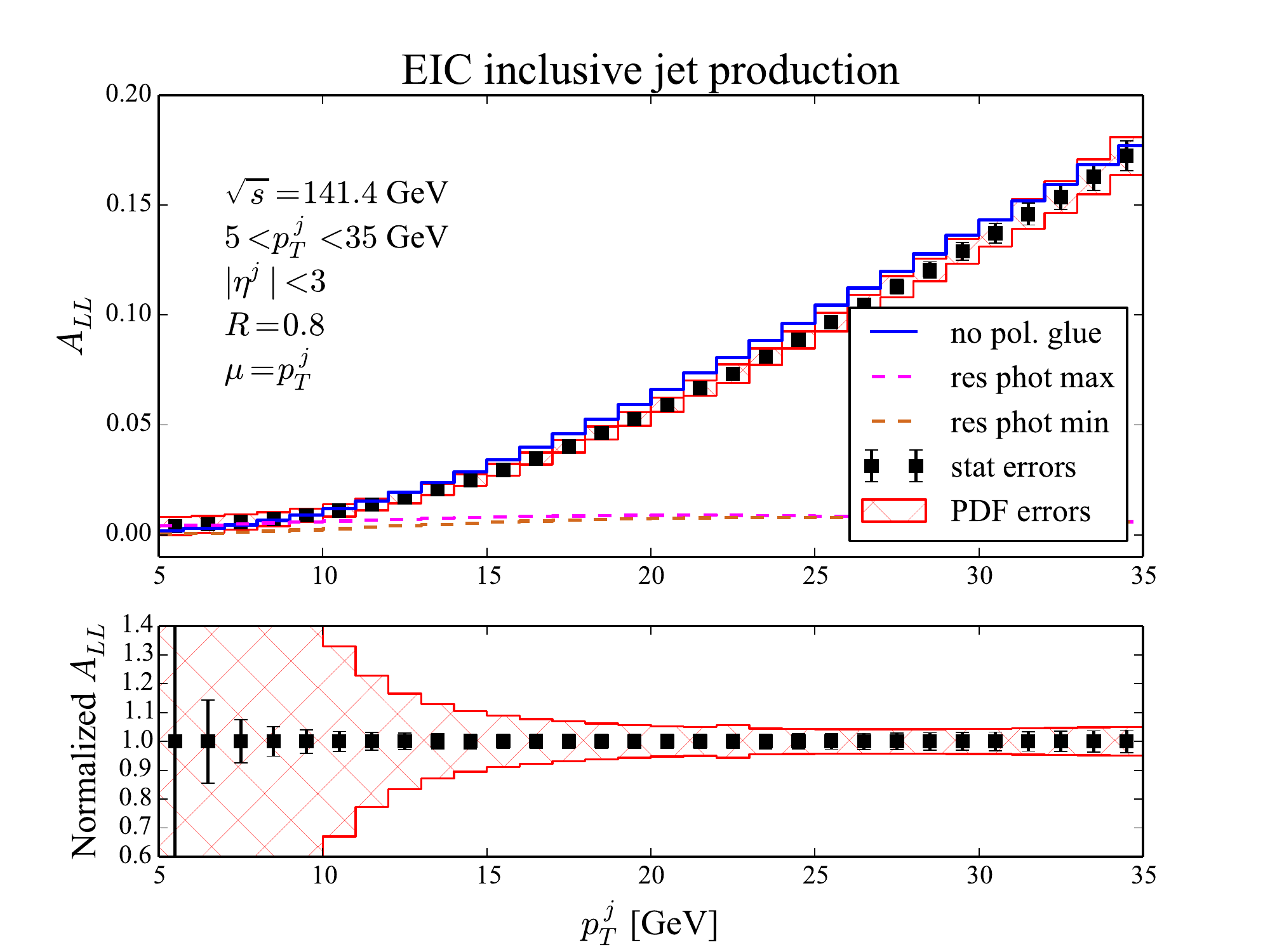, width=3.3in}
\hskip 0.2in
\psfig{file=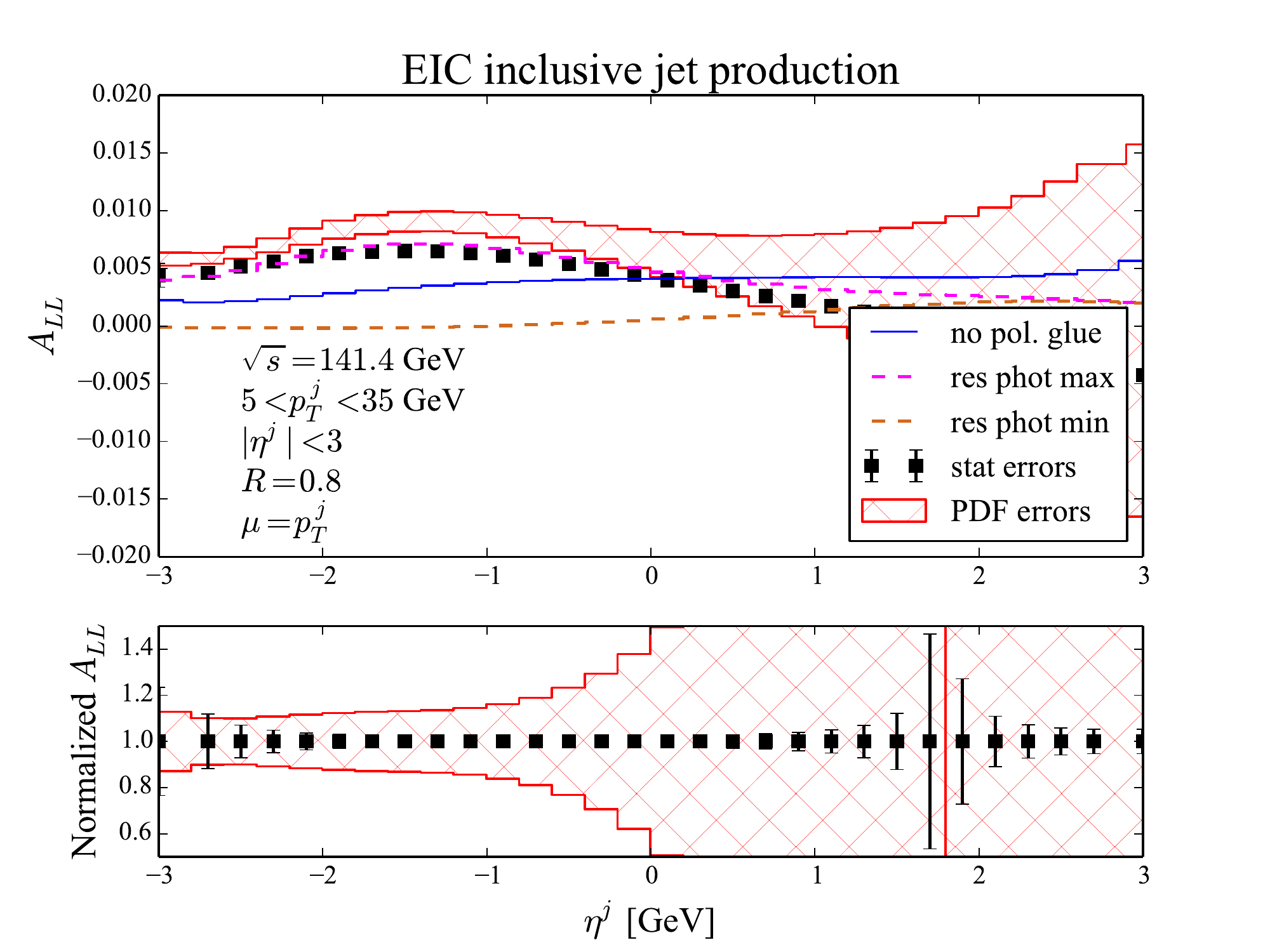, width=3.3in}
\caption{Spin asymmetry as a function of jet transverse momentum (left panel) and jet pseudorapidity (right panel). The resolved photon contribution is shown separately in the upper panel of each plot.  The lower panels normalize the results to the central values in order to more clearly illustrate the errors.}
\label{fig:sqrts141-ALL}
\eef

To illustrate the structure of the asymmetry we split it into partonic contributions in Fig.~\ref{fig:sqrts141-ALL-split}.  In these plots we have kept only the indicated partonic channel in the numerator of the asymmetry, but all channels in the denominator.  This makes each term directly proportional to the desired polarized distribution functions.  The dominant contribution to the asymmetry at intermediate-to-high $p_T^j$ comes from the $ql$ channel.  At intermediate $p_T^j$ the $q\gamma$ and $g\gamma$ contributions are important.  It is interesting to note that the sensitivity to the polarized gluon distribution comes from the $g\gamma$ channel, which occurs only for inclusive jet production with $Q^2 \approx 0$ without a tagged lepton.  As we will see explicitly later the sensitivity to $\Delta f_{g/H}$ vanishes upon imposing a large cut on $Q^2$.  This demonstrates the importance of jet production measurements without a tagged lepton at a future EIC to give a direct determination of the polarized gluon distribution.  The $A_{LL}$ distribution is dominated at low $\eta^j$ by the resolved photon term in the minimal model of this distribution, while at high-$\eta^j$ the $g\gamma$ channel determines the shape of the distribution.  The resolved photon contribution is small if instead the maximal model of Ref.~\cite{Gluck:1992fy} is assumed. Again, this contribution only occurs for inclusive jet production without a tagged lepton.

\begin{figure}[h]
\psfig{file=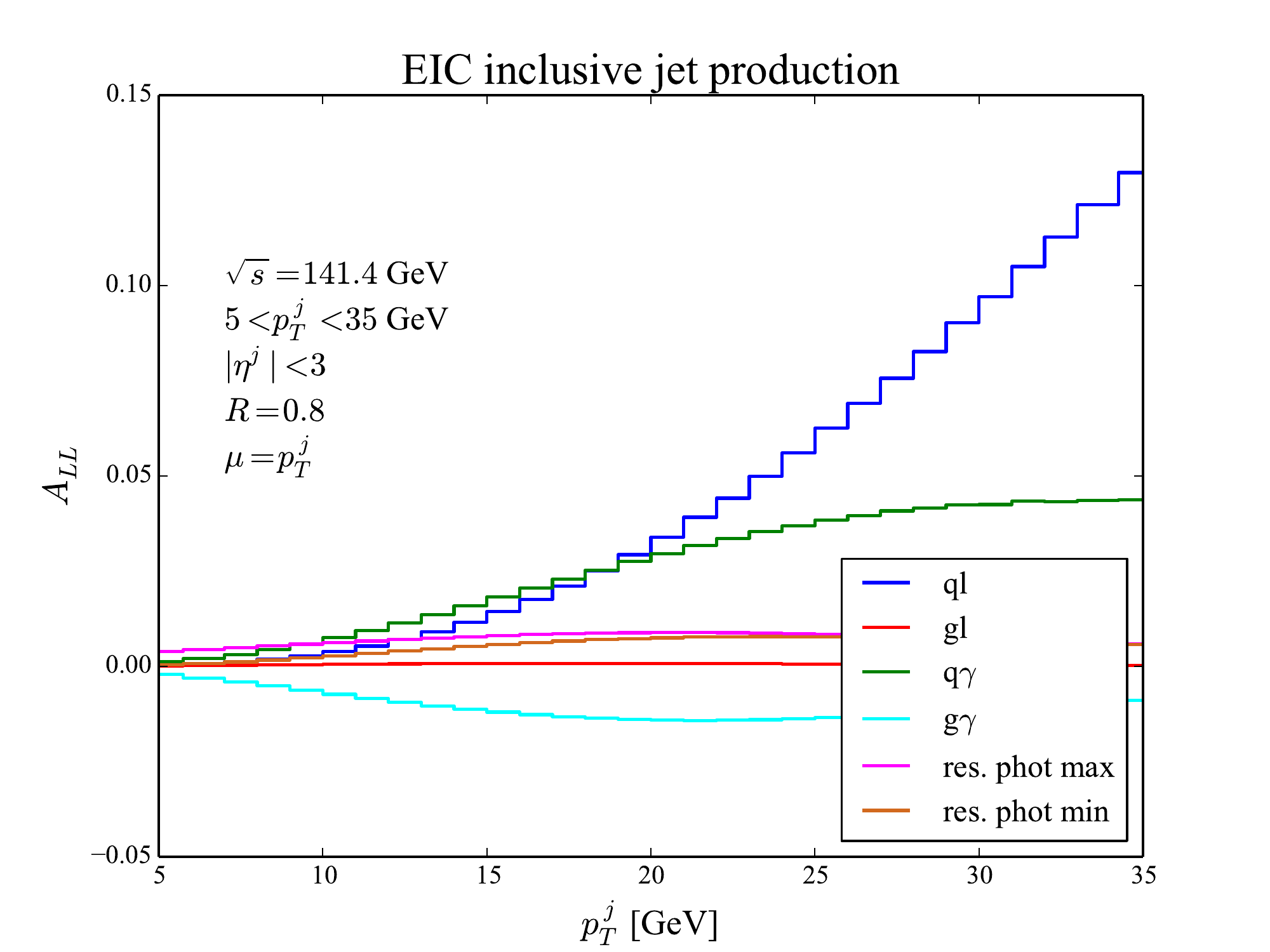, width=3.3in}
\hskip 0.2in
\psfig{file=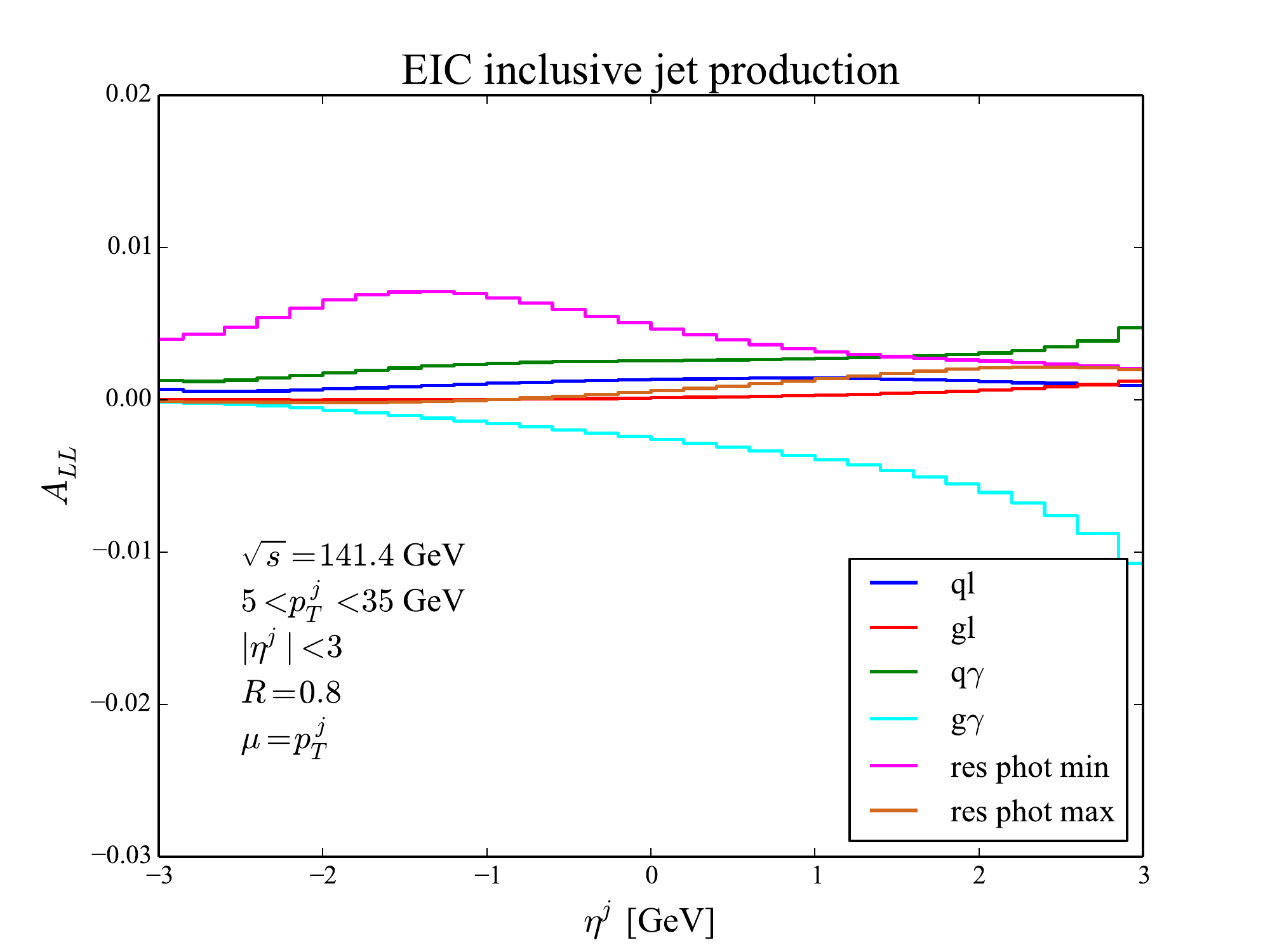, width=3.3in}
\caption{Splits of the spin asymmetry as functions of the jet transverse momentum and pseudorapidity into partonic channels.}
\label{fig:sqrts141-ALL-split}
\eef

In order to increase the sensitivity of $A_{LL}$ to the polarized gluon distribution, the previous results motivate isolating the high-$p_T^j$ region where Fig.~\ref{fig:sqrts141-ALL} indicates that the contribution of this quantity becomes significant.  We impose the more stringent cut $p_T^j > 20$ GeV and show both the $A_{LL}$ distribution and its split into partonic channels in Fig.~\ref{fig:sqrts141-ALL-pt20}.  We note that the result without the polarized gluon differs significantly for positive $\eta^j$ from the one with the polarized gluon included.  Both models for the polarized distribution of the photon are small for this $p_T^j$ cut, suggesting excellent sensitivity of this quantity to the polarized structure of the proton.  The split into partonic channels shows that the dominant channels are the $ql$, $q\gamma$, and $g\gamma$ ones.  We note a significant cancellation between the $q\gamma$ and $g\gamma$ channels that is relaxed when $\Delta f_{g/H}$ is turned off, leading to the larger asymmetry without the polarized gluon in the left plot of Fig.~\ref{fig:sqrts141-ALL-pt20}.

\begin{figure}[h]
\psfig{file=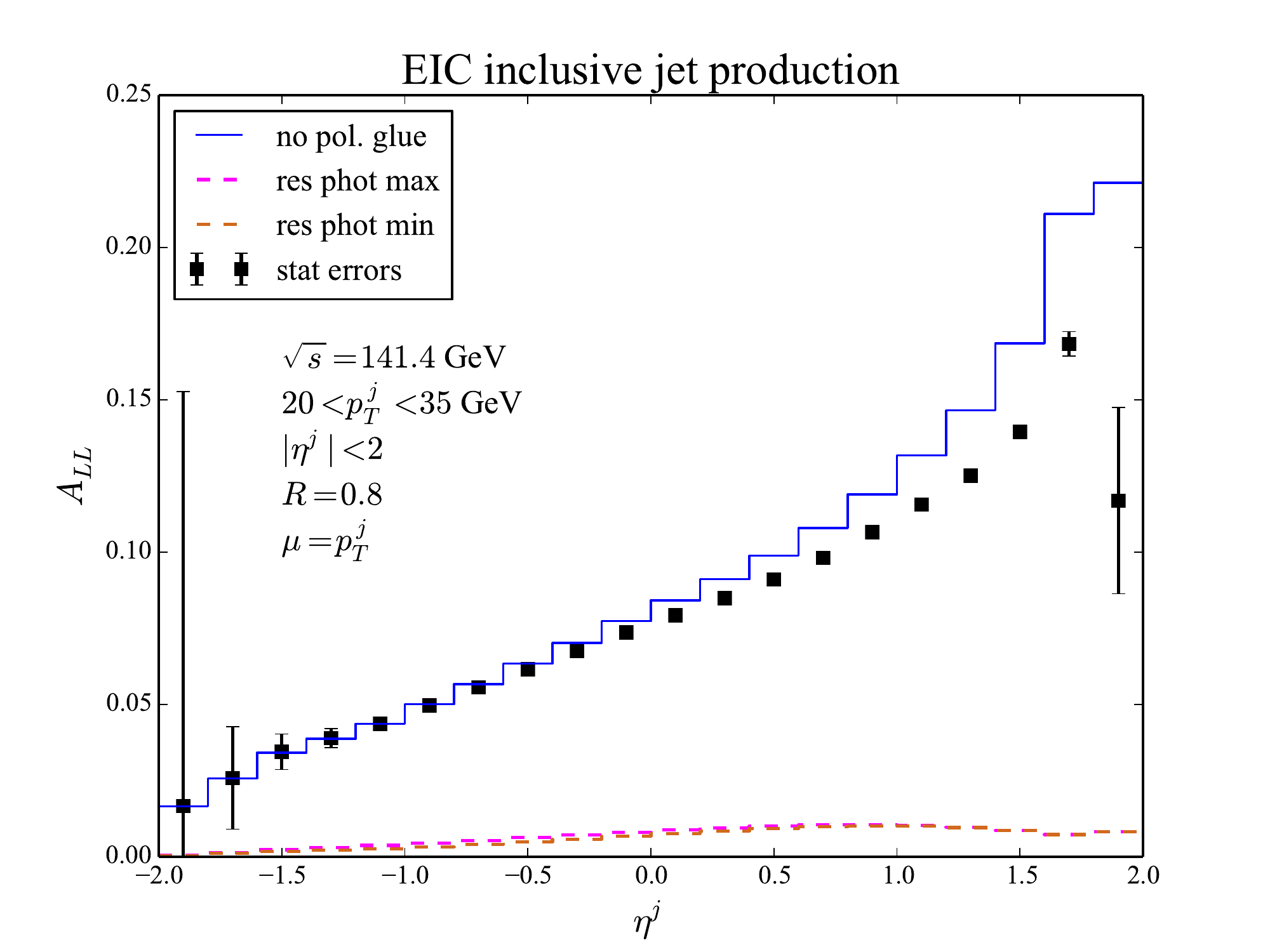, width=3.3in}
\hskip 0.2in
\psfig{file=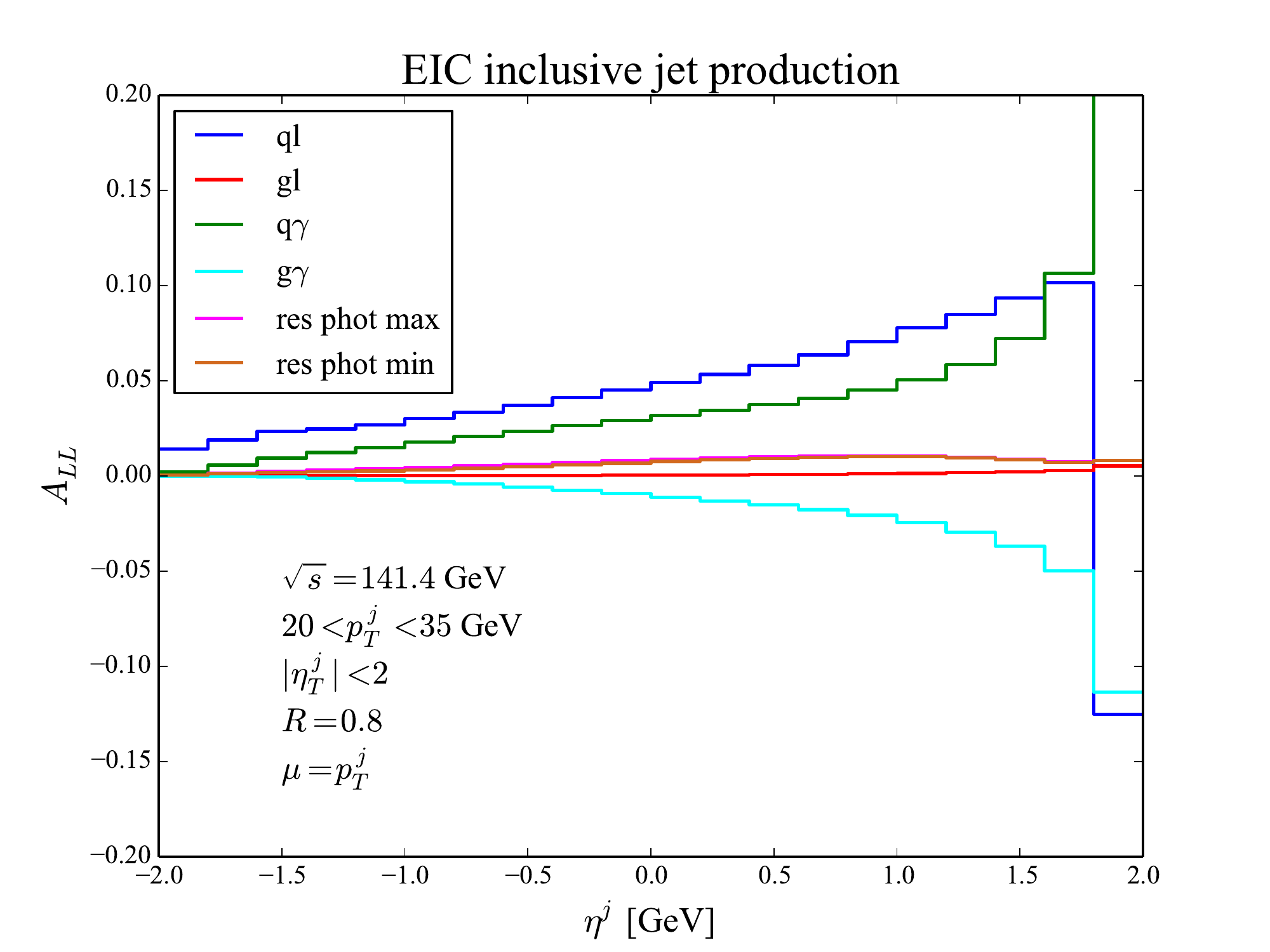, width=3.3in}
\caption{Spin asymmetry as a function of $\eta^j$ with the cut $p_T^j > 20$ GeV.}
\label{fig:sqrts141-ALL-pt20}
\eef

Since the NLO real-emission corrections to the $ql$ and $gl$ channels contain two final-state partons that can be arbitrarily separated in azimuthal angle and pseudorapidity due to the presence of the wide-angle lepton, as can be seen from Fig.~\ref{fig:NLOlepdiag}, there is a non-trivial dependence on the anti-$k_T$ radius parameter $R$ that begins at this order.  In order to investigate the dependence on $R$ we study both the unpolarized cross section and $A_{LL}$ for the two choices $R=0.2$ and $R=0.8$ in Figs.~\ref{fig:sqrts141-totcr-Rcomp} and~\ref{fig:sqrts141-ALL-Rcomp}.  We see that the dependence on the jet radius is minimal.  The choice $R=0.8$ gives a slightly smaller cross section and a softer asymmetry as a function of $p_T^j$, but the effects on $A_{LL}$ are well within other theoretical uncertainties and the expected experimental errors.

\begin{figure}[h]
\psfig{file=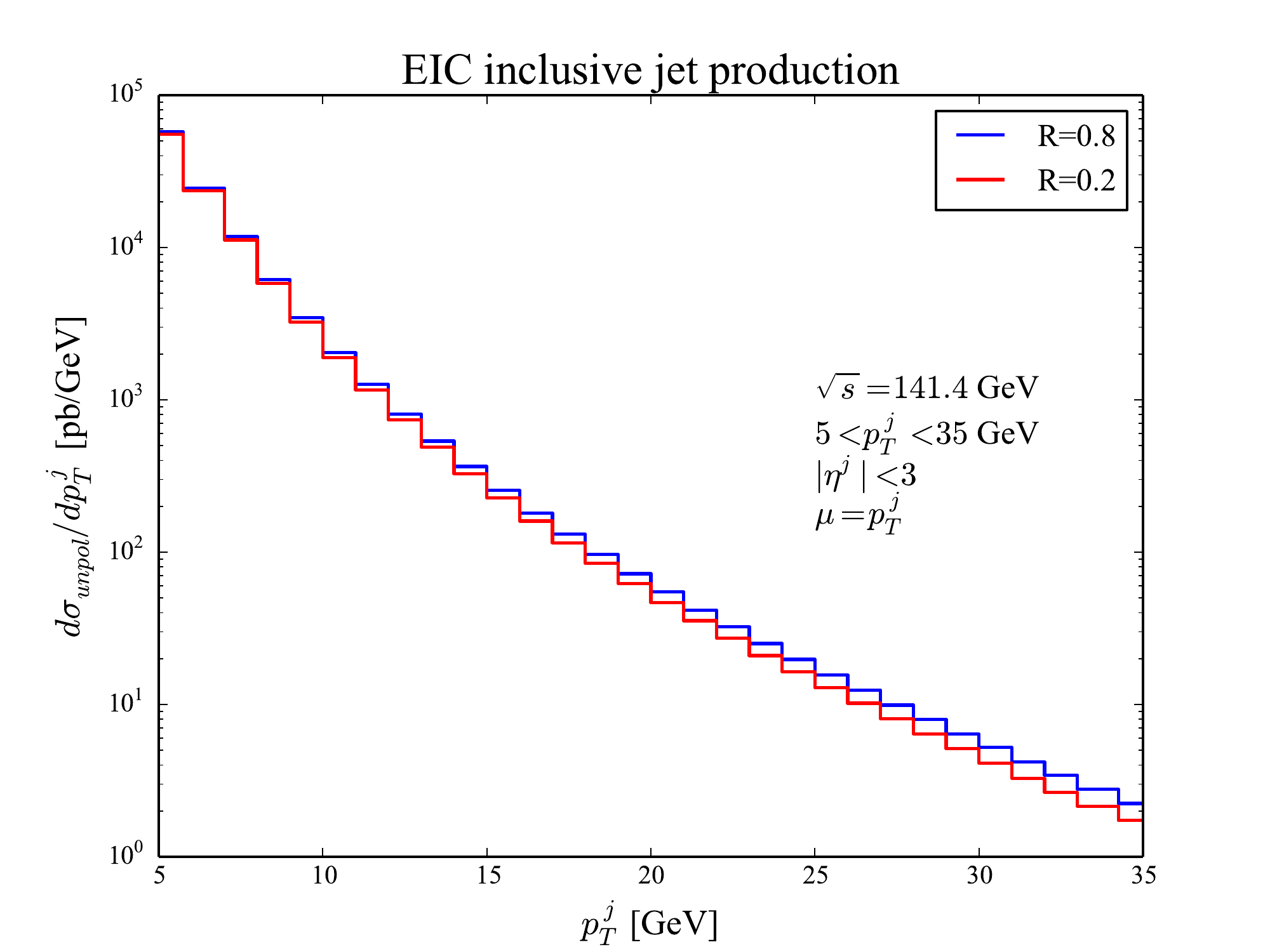, width=3.3in}
\hskip 0.2in
\psfig{file=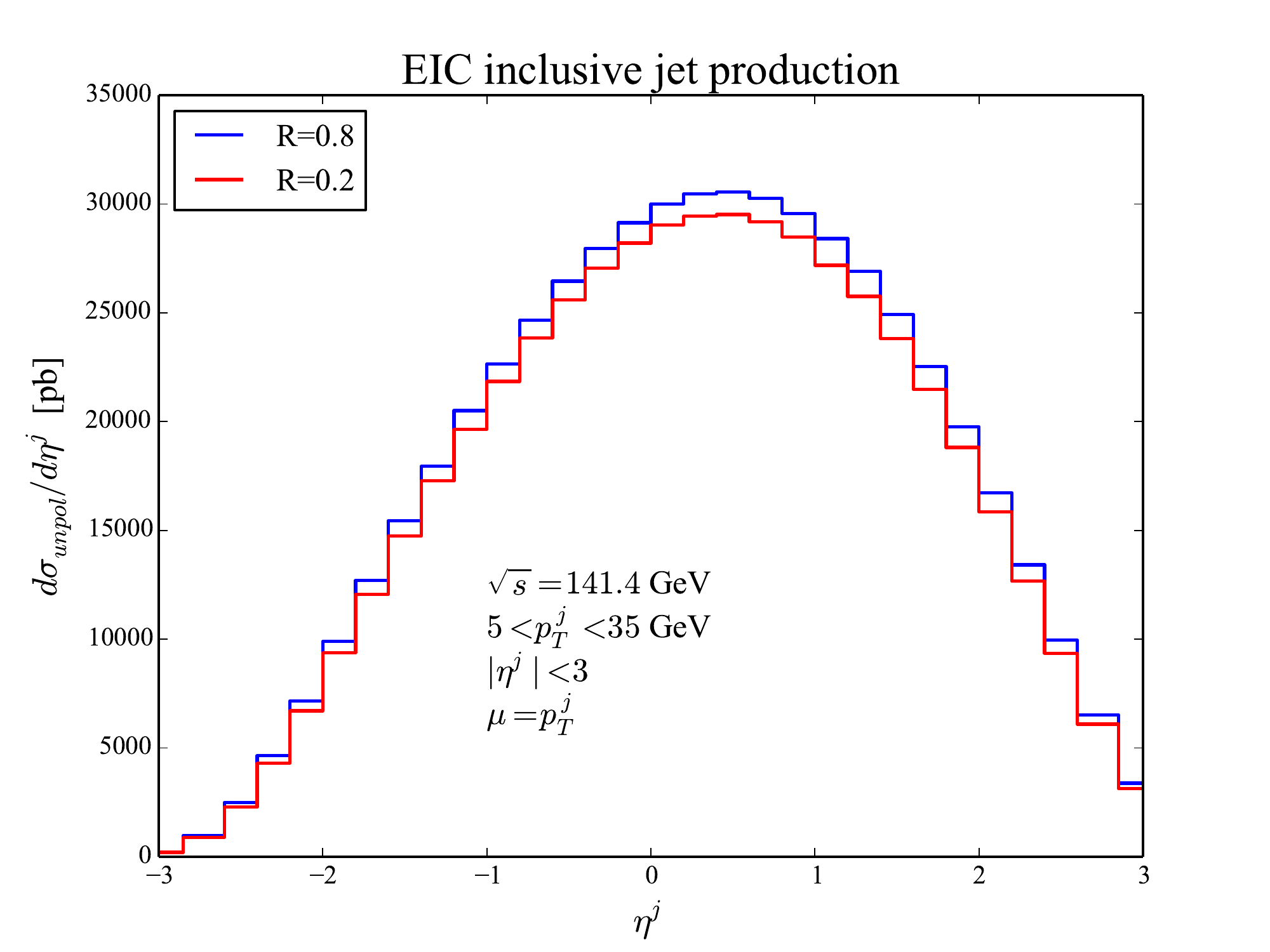, width=3.3in}
\caption{Unpolarized cross section as a function of jet transverse momentum (left panel) and jet pseudorapidity (right panel) for the two jet radius parameters $R=0.2$ and $R=0.8$.}
\label{fig:sqrts141-totcr-Rcomp}
\eef

\begin{figure}[h]
\psfig{file=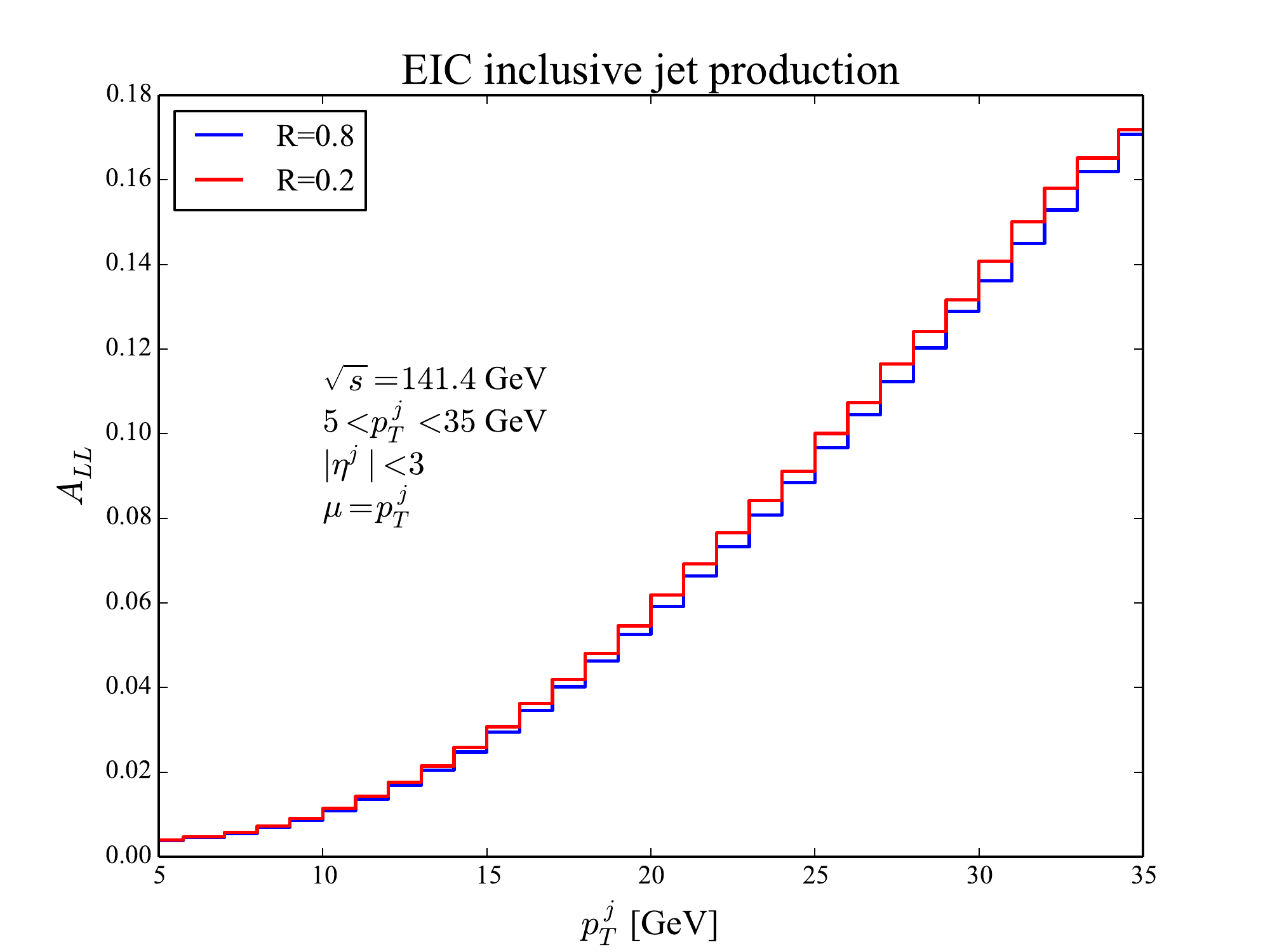, width=3.3in}
\hskip 0.2in
\psfig{file=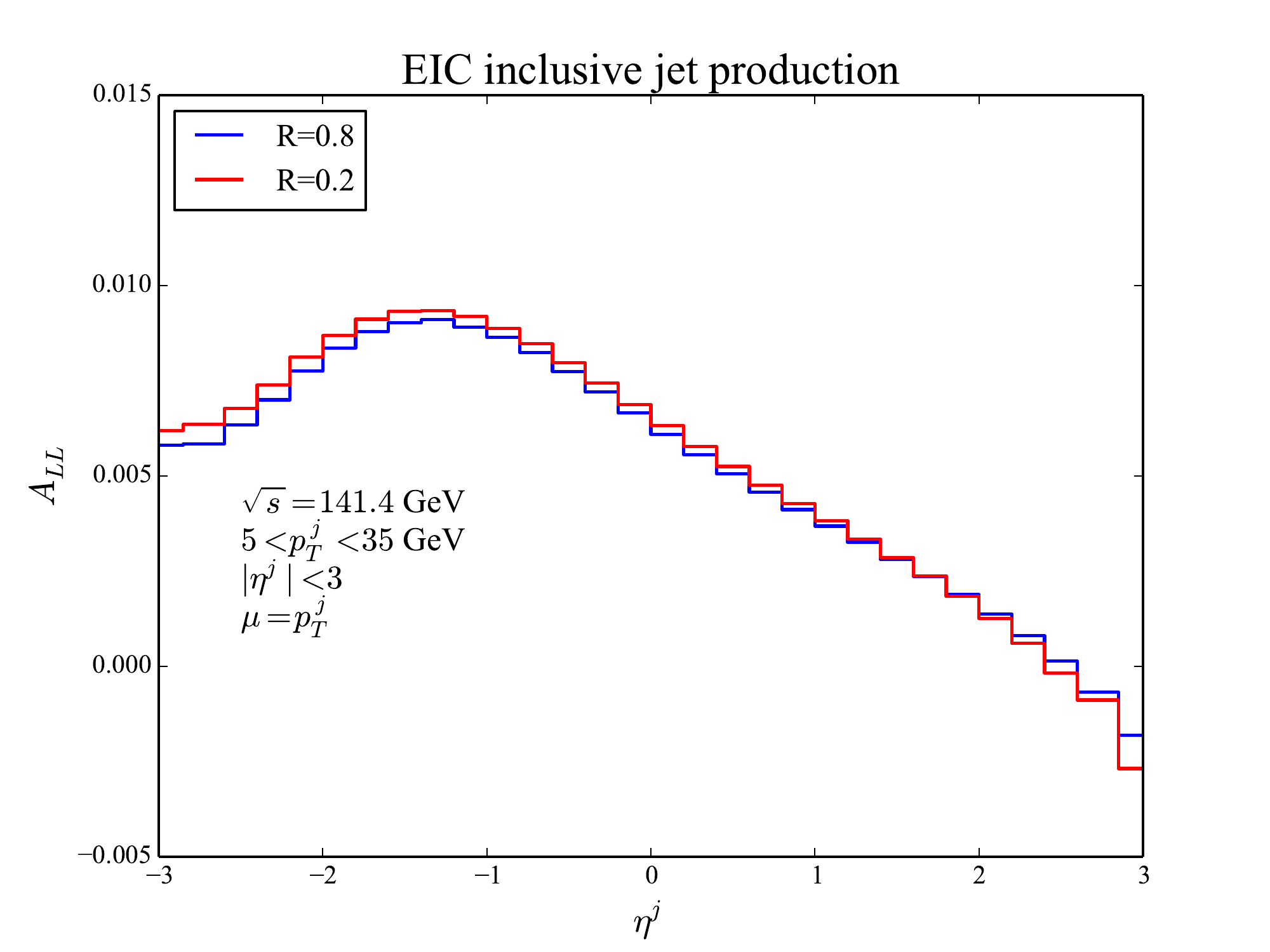, width=3.3in}
\caption{Spin asymmetry as a function of jet transverse momentum (left panel) and jet pseudorapidity (right panel) for the two jet radius parameters $R=0.2$ and $R=0.8$.}
\label{fig:sqrts141-ALL-Rcomp}
\eef

Finally, although we have used the NNPDF 1.1 polarized distribution functions for the presented results, other parameterizations of these quantities are available in the literature.  To test the sensitivity of the EIC to different PDF parameterizations we compare the $A_{LL}$ results obtained using NNPDF to those obtained using the DSSV global fit to the available data~\cite{deFlorian:2014yva}.  The $A_{LL}$ distribution as a function of both $p_T^j$ and $\eta^j$ is shown in Fig.~\ref{fig:sqrts141-ALL-DSSV} for the central values of both PDF sets.  There are slight differences between the asymmetries obtained using the two different parameterizations.  However, comparison with Fig.~\ref{fig:sqrts141-ALL} reveals that all differences are well within the PDF uncertainties as estimated in the NNPDF fit.

\begin{figure}[h]
\psfig{file=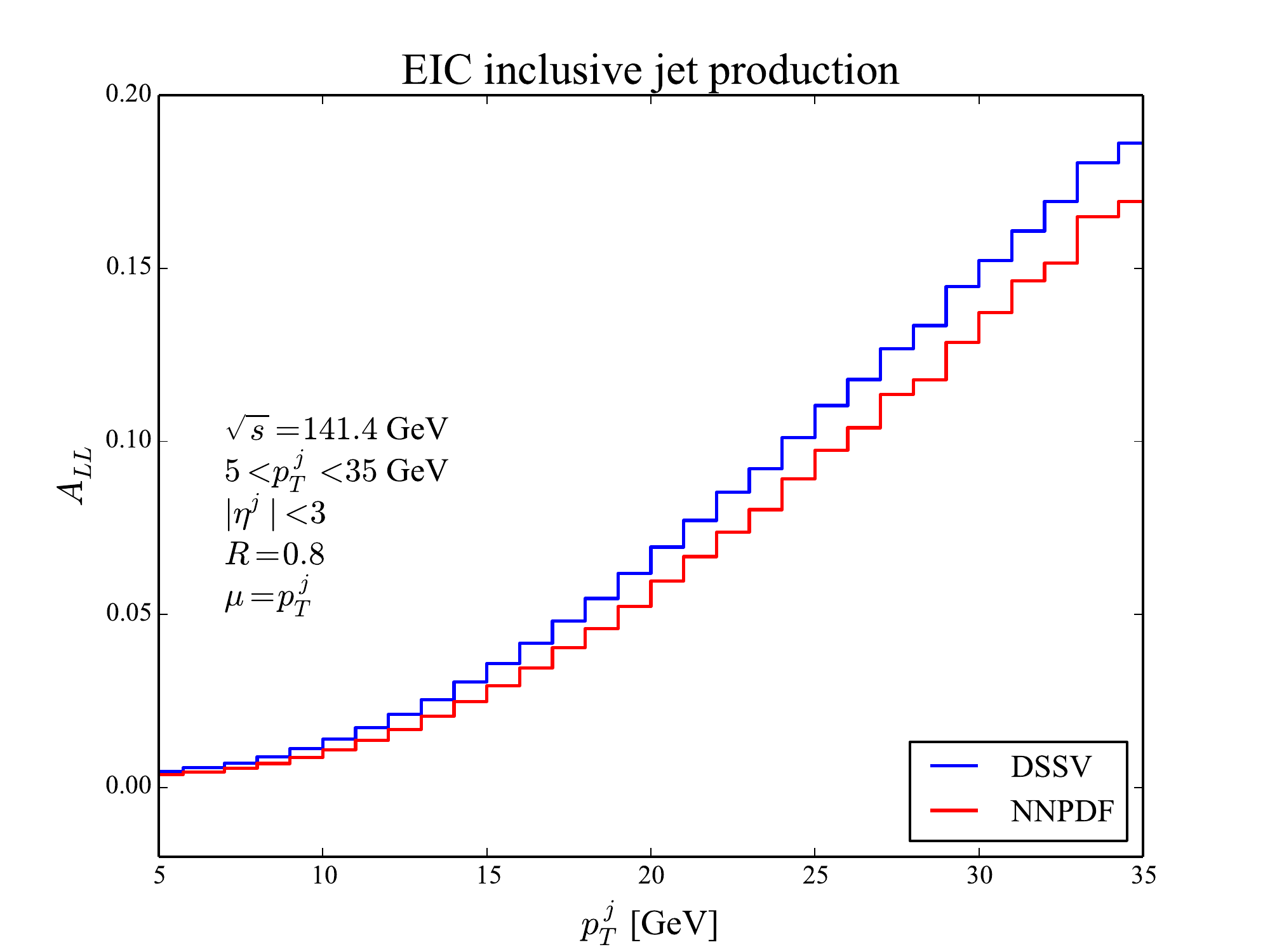, width=3.3in}
\hskip 0.2in
\psfig{file=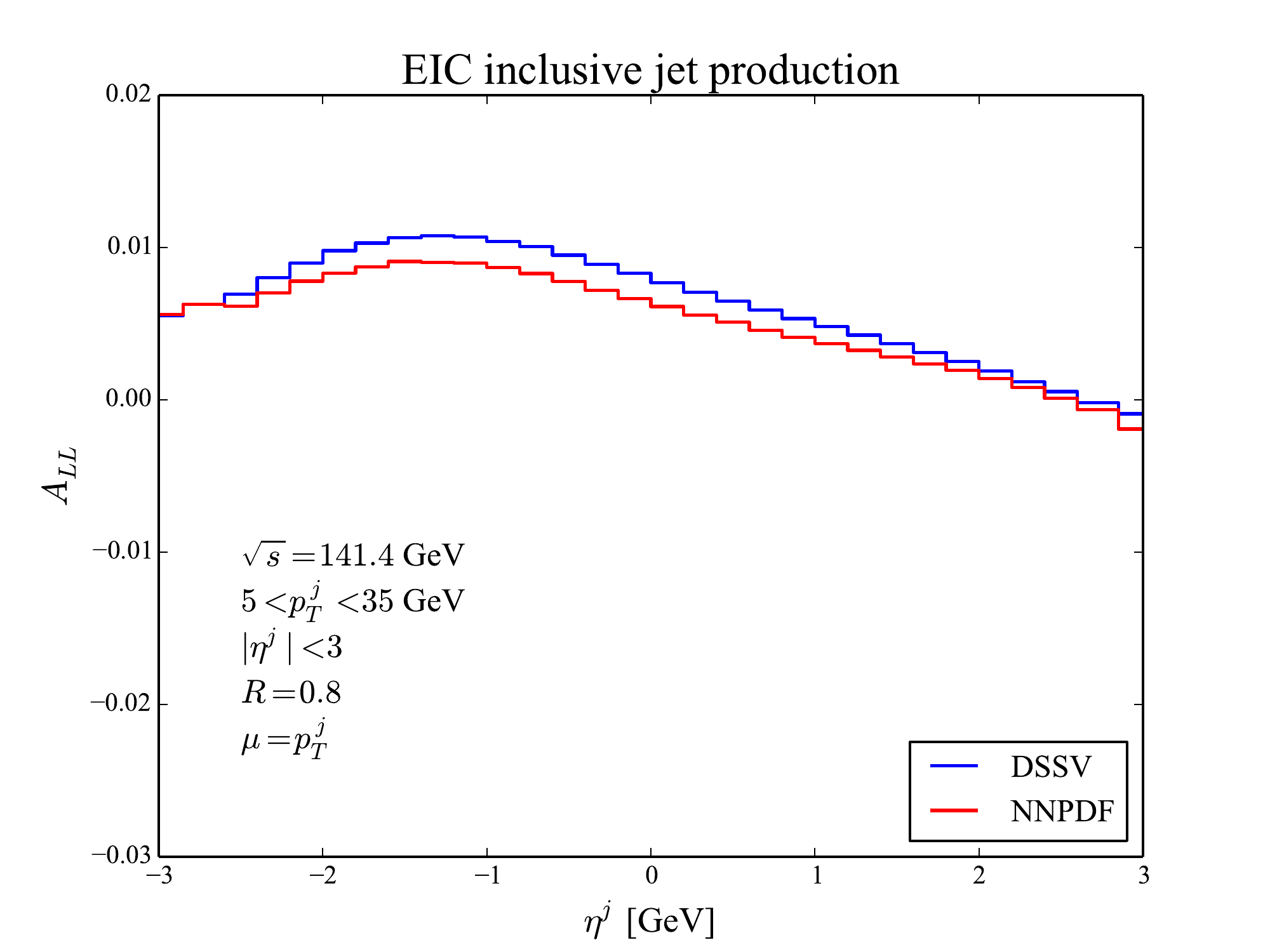, width=3.3in}
\caption{Spin asymmetry as a function of jet transverse momentum (left panel) and jet pseudorapidity (right panel) for the NNPDF and DSSV extractions of polarized PDFs.}
\label{fig:sqrts141-ALL-DSSV}
\eef

\subsection{Jet production with a tagged lepton}

We next study the spin asymmetry when the cut $Q^2 > 10$ GeV$^2$ is imposed, indicating the presence of a wide-angle lepton in the final-state.  As mentioned previously this cut significantly changes the structure of the cross section.  The virtual photon emitted from the lepton is far off-shell when this cut is imposed.  Since the virtuality is much larger than the electron mass there are no longer contributions from the $q\gamma$ and $g\gamma$ channels in our theoretical framework, since these contributions become a useful description only when $Q^2  \lesssim m_{elec}^2$.  Similarly, since $Q^2 \gg \Lambda_{QCD}^2$ there are no longer resolved photon contributions.  Splitting of the photon into partons at large $Q^2$ happens perturbatively without large logarithmic contributions at higher orders in $\alpha_s$ than those considered in this study, and are not enhanced.  This leaves only the $ql$ and $gl$ channels to consider.  

The results for the distributions of $A_{LL}$ in $p_T^j$ and $\eta^j$ are shown in Fig.~\ref{fig:sqrts141-ALL-Q2cut}, while the splits into the $ql$ and $gl$ channels is shown in Fig.~\ref{fig:sqrts141-ALL-Q2cut-split}.  We note that the PDF uncertainties are smaller than for inclusive jet production, particularly at low-$p_T^j$ and high-$\eta^j$.  These regions receive significant contributions from the $g\gamma$ channel in the inclusive jet production case but not here, indicating the reason for this difference.  The $ql$ channel dominates for all values of transverse momentum, and for all but the very forward region of pseudorapidity.  

\begin{figure}[h]
\psfig{file=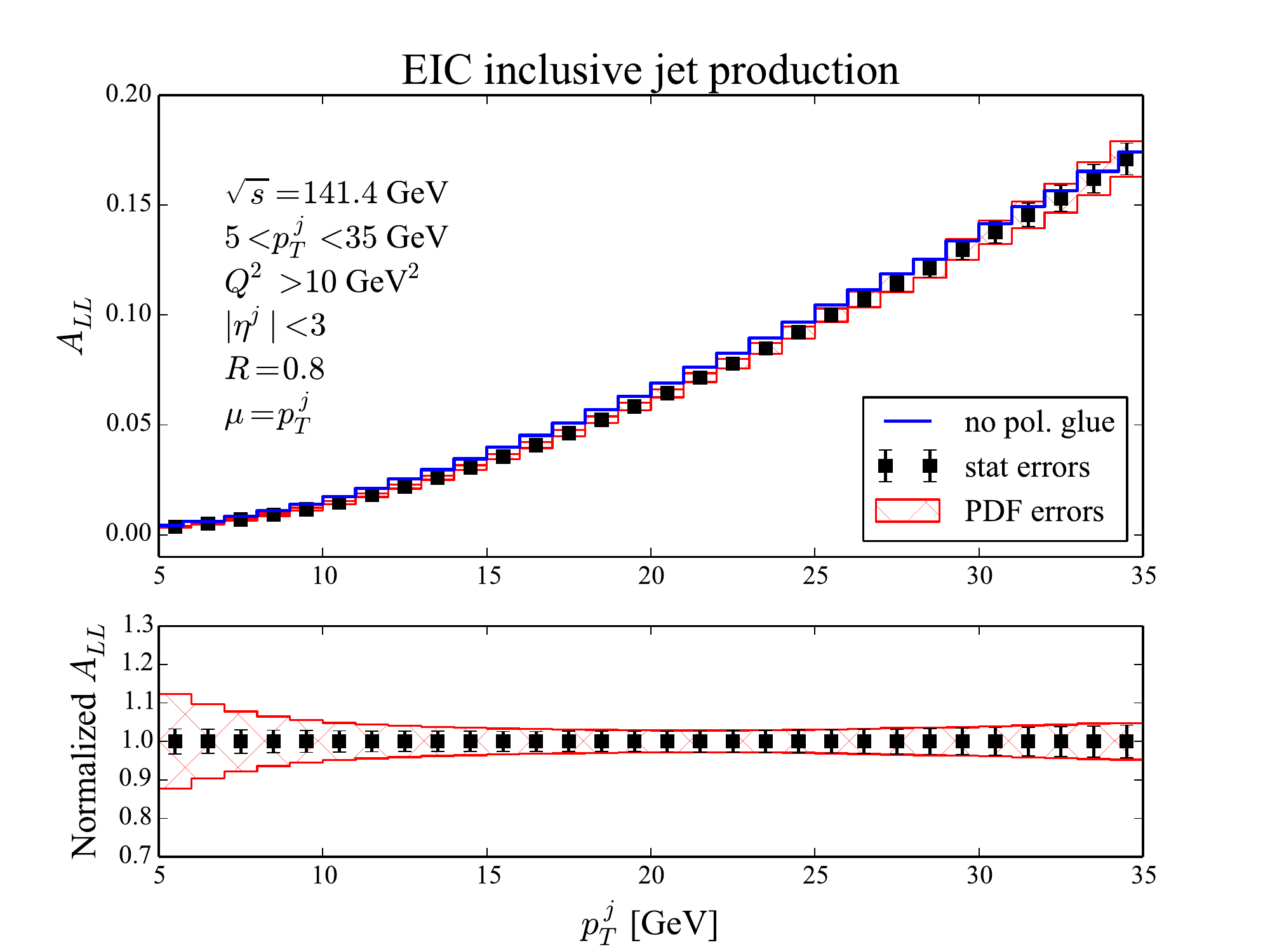, width=3.3in}
\hskip 0.2in
\psfig{file=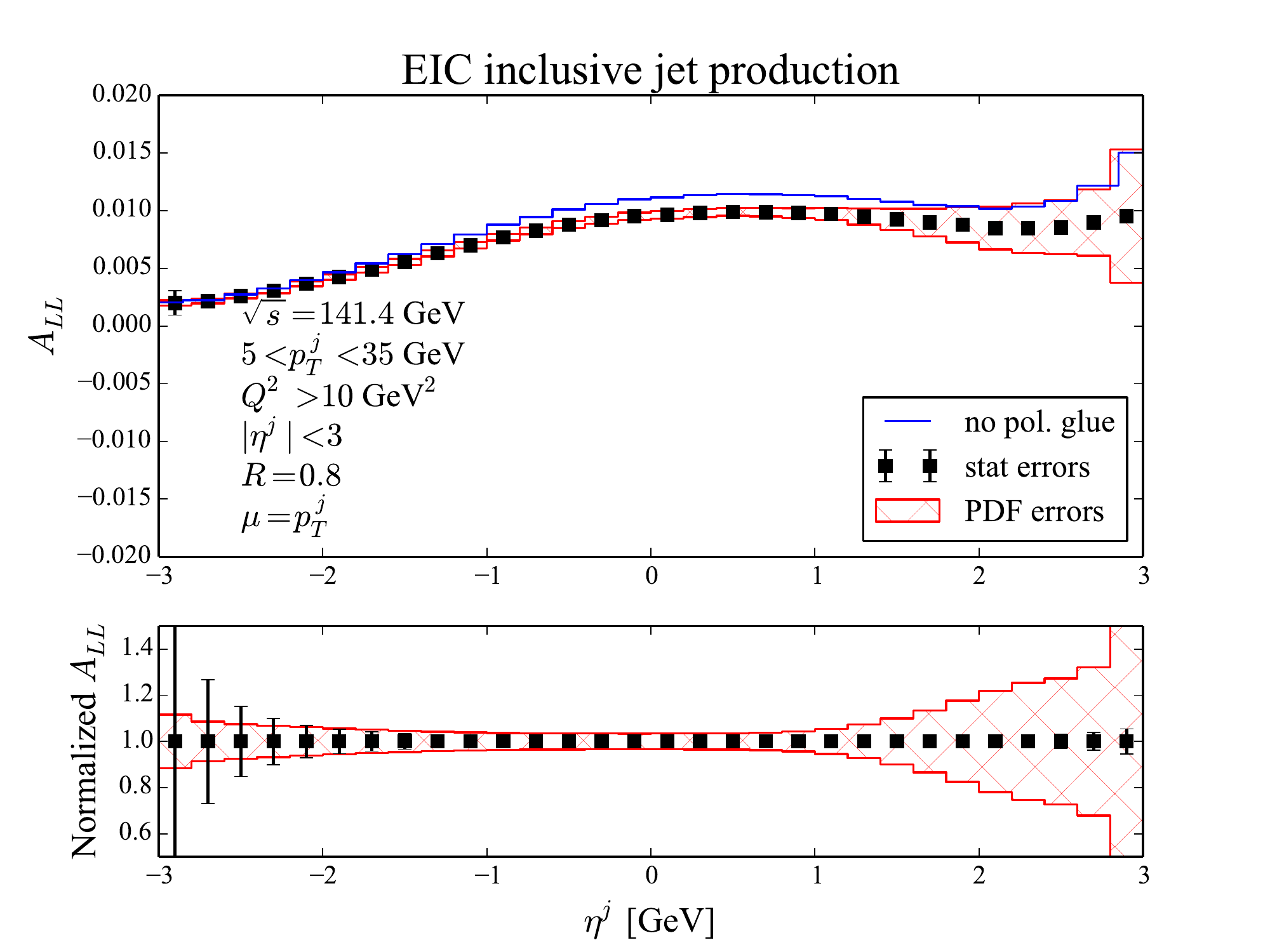, width=3.3in}
\caption{Spin asymmetry as a function of jet transverse momentum (left panel) and jet pseudorapidity (right panel). The resolved photon contribution is shown separately in the upper panel of each plot.  The lower panels normalize the results to the central values in order to more clearly illustrate the errors.  The cut $Q^2 > 10$ GeV$^2$ has been imposed.}
\label{fig:sqrts141-ALL-Q2cut}
\eef

\begin{figure}[h]
\psfig{file=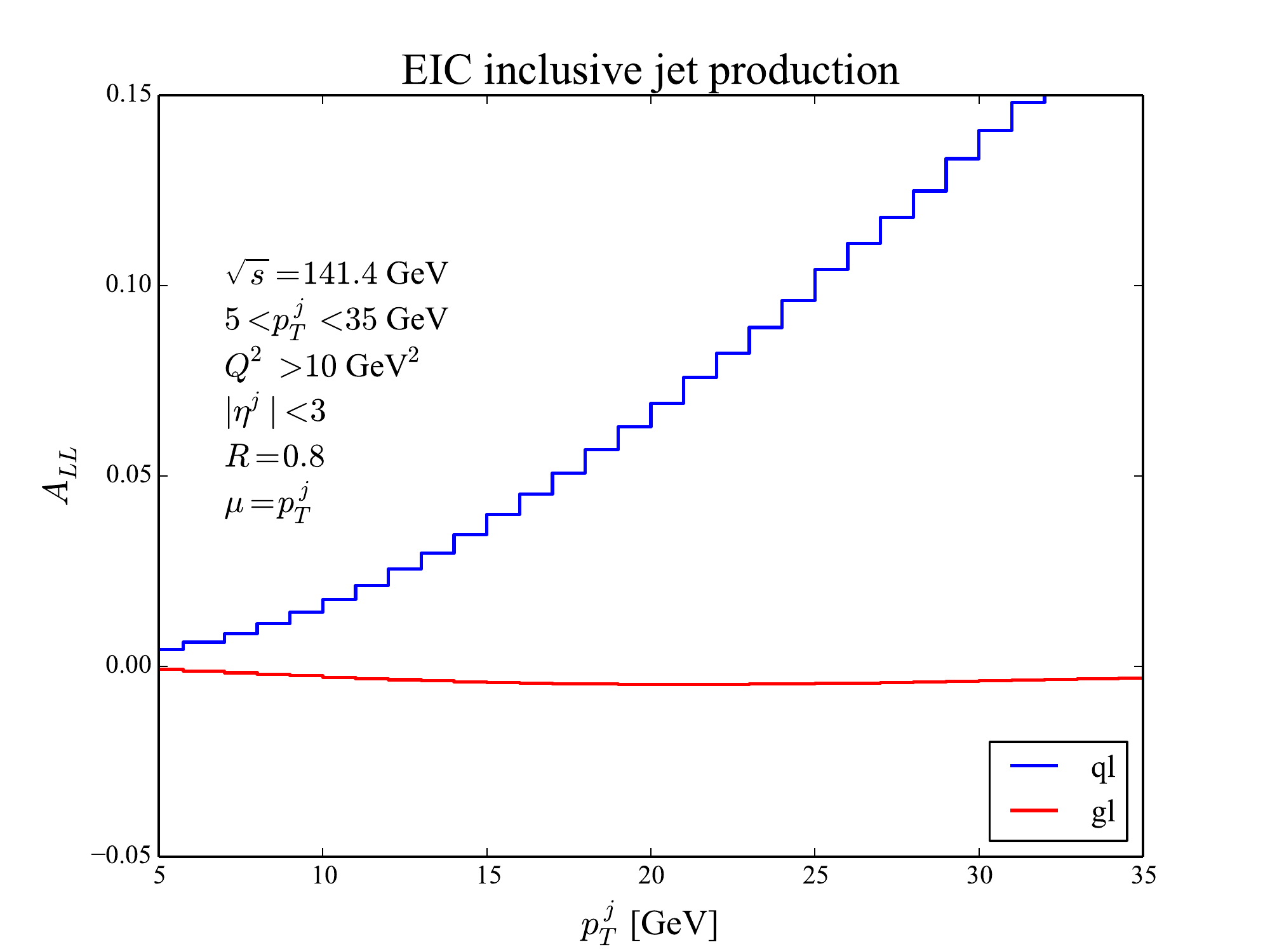, width=3.3in}
\hskip 0.2in
\psfig{file=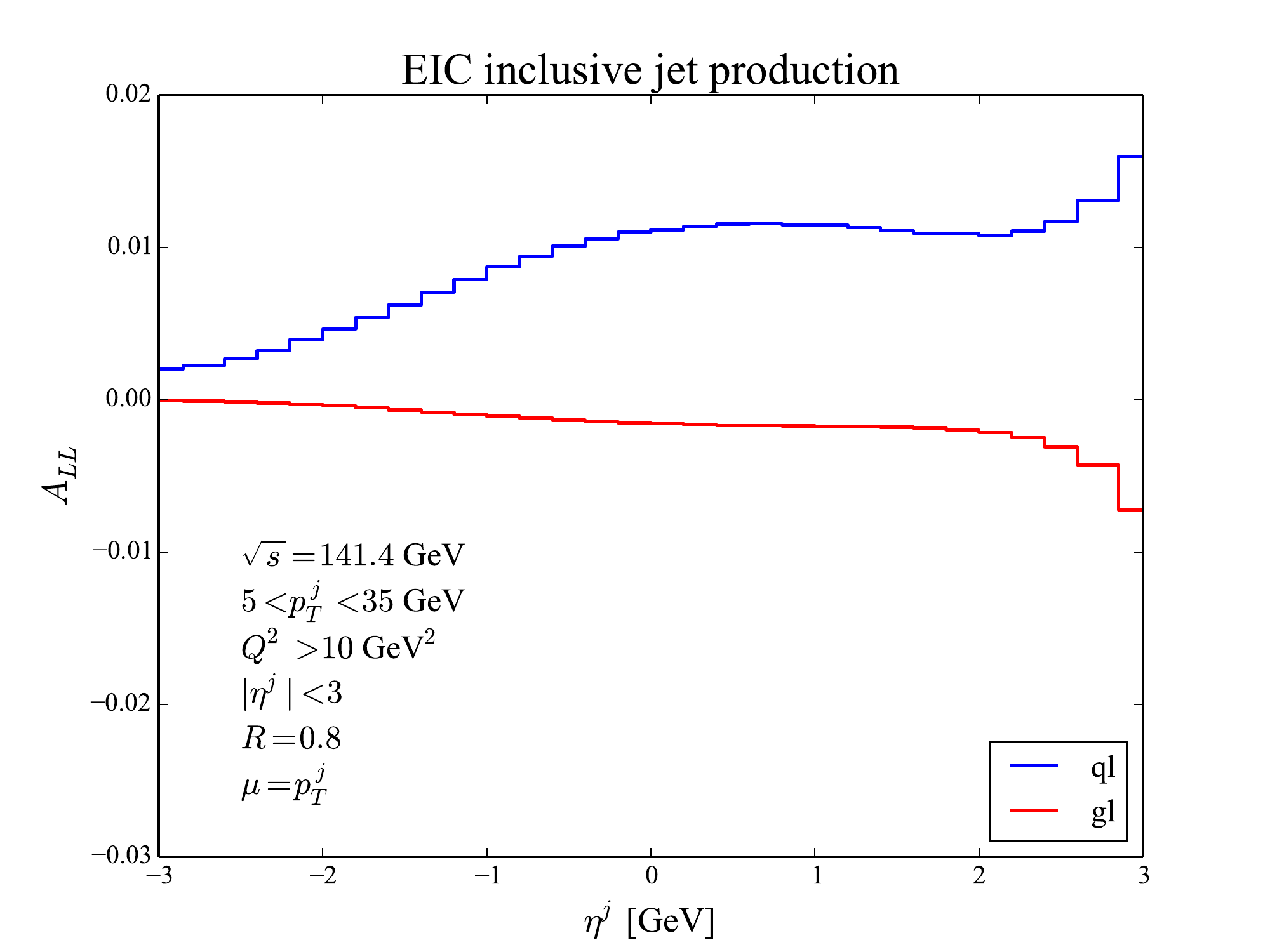, width=3.3in}
\caption{Splits of the spin asymmetry as functions of the jet transverse momentum and pseudorapidity into partonic channels.  The cut $Q^2 > 10$ GeV$^2$ has been imposed.}
\label{fig:sqrts141-ALL-Q2cut-split}
\eef

\section{Results for $\sqrt{s} = 63.2$ GeV}
\label{sec:63}

We now study jet production at an EIC with center-of-mass energy $\sqrt{s}=63.2$ GeV.  Due to the reduced collision energy the accessible kinematic range of the jets is reduced, and we consequently focus on the phase-space region $5 < p_T^j< 30$ GeV and $|\eta^j|<2.5$.  The collisions at this energy typically occur at higher Bjorken-$x$, and this determines much of the observed phenomenology.

We show first in Fig.~\ref{fig:sqrts63-totcr} the unpolarized cross section as a function of both $p_T^j$ and $\eta^j$.  The corresponding split into partonic channels is shown in Fig.~\ref{fig:sqrts63-totcr-split}.  Although still the largest partonic channel at low $p_T^j$, the resolved photon contribution is no longer a factor of several larger than the other channels, unlike for $\sqrt{s}=141.4$ GeV.  This is because the parton-in-lepton distributions required to obtain these channels fall off more rapidly with Bjorken-$x$ due to the multiple collinear splittings needed to generate them.  This can be observed from the additional convolution in Eq.~(\ref{eq:partinlep}).  Above $p_T^j \approx 10$ GeV the $ql$ channel dominates, with the second largest being the $q\gamma$ channel.  All channels contribute non-negligibly to the $\eta^j$ distribution except $gl$, which is small throughout phase space.  We note that the $gl$ channel is small and negative for small $p_T^j$, which explains why it  begins at $p_T^j \approx 25$ GeV in the left panel of Fig.~\ref{fig:sqrts63-totcr-split}.  This channel by itself is not a physical observable, but is only a component of the full NLO cross section defined in the $\overline{\text{MS}}$ scheme.  We note that the PDF errors are larger at high $p_T^j$ than the corresponding errors for $\sqrt{s}=141.4$ GeV.  Ths is because the PDFs are being probed at very high Bjorken-$x$ where constraints from current data are limited.  However, the estimated statistical errors also become large in this region.  

\begin{figure}[h]
\psfig{file=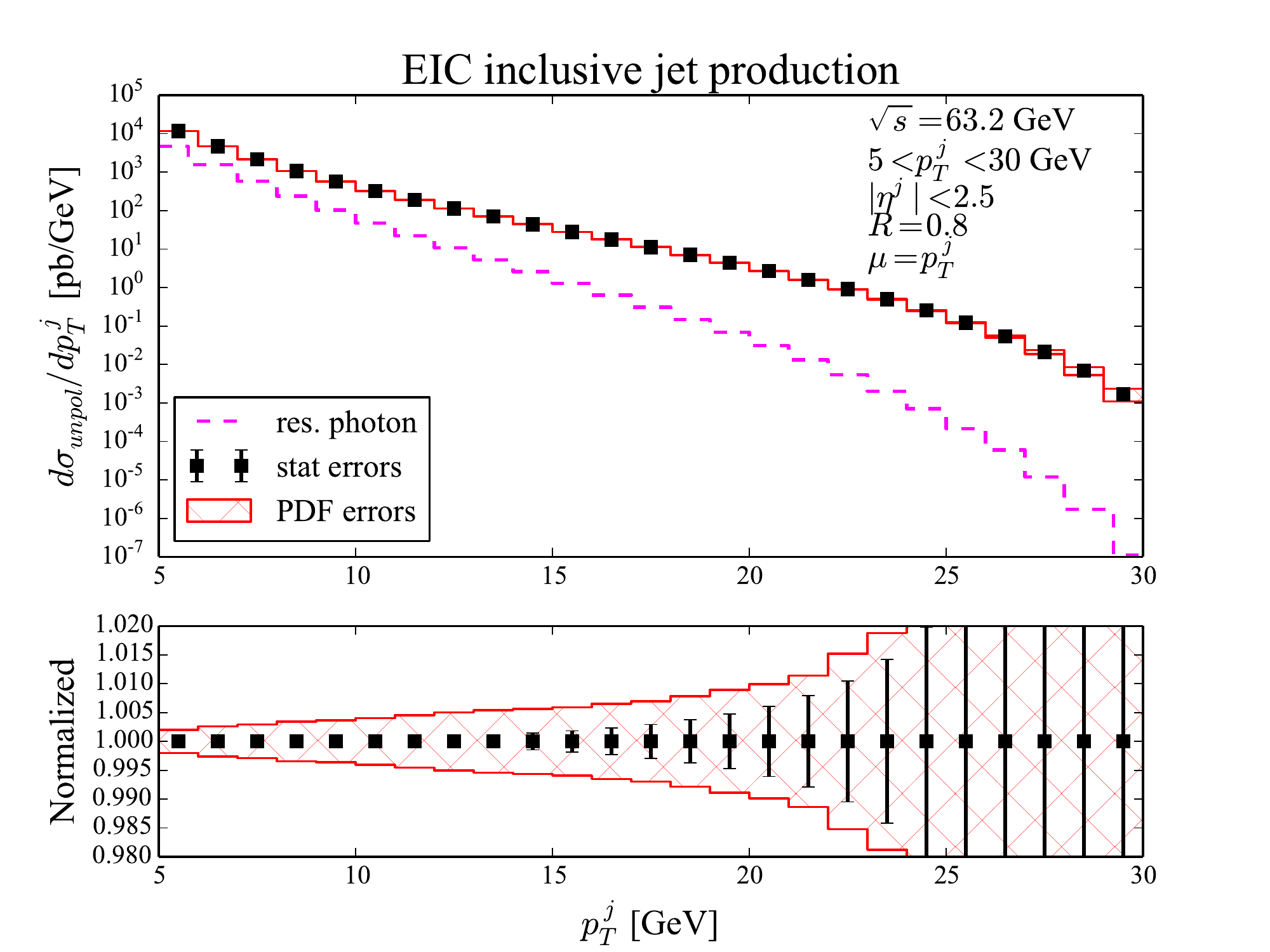, width=3.3in}
\hskip 0.2in
\psfig{file=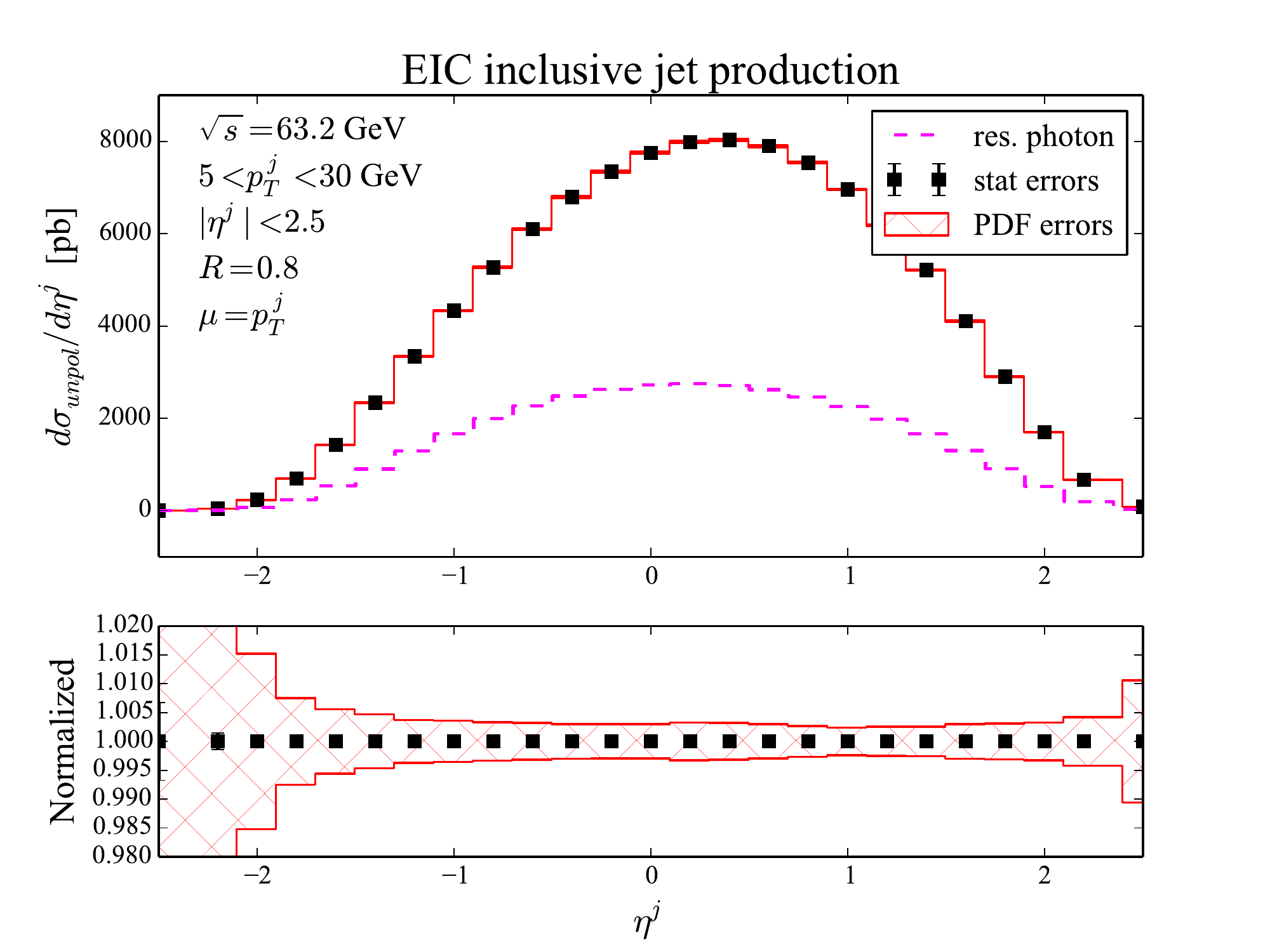, width=3.3in}
\caption{Total unpolarized cross section as a function of jet transverse momentum (left panel) and jet pseudorapidity (right panel) for $\sqrt{s}=63.2$ GeV. The resolved photon contribution is shown separately in the upper panel of each plot.  The lower panels normalize the results to the central values in order to more clearly illustrate the errors.}
\label{fig:sqrts63-totcr}
\eef

\begin{figure}[h]
\psfig{file=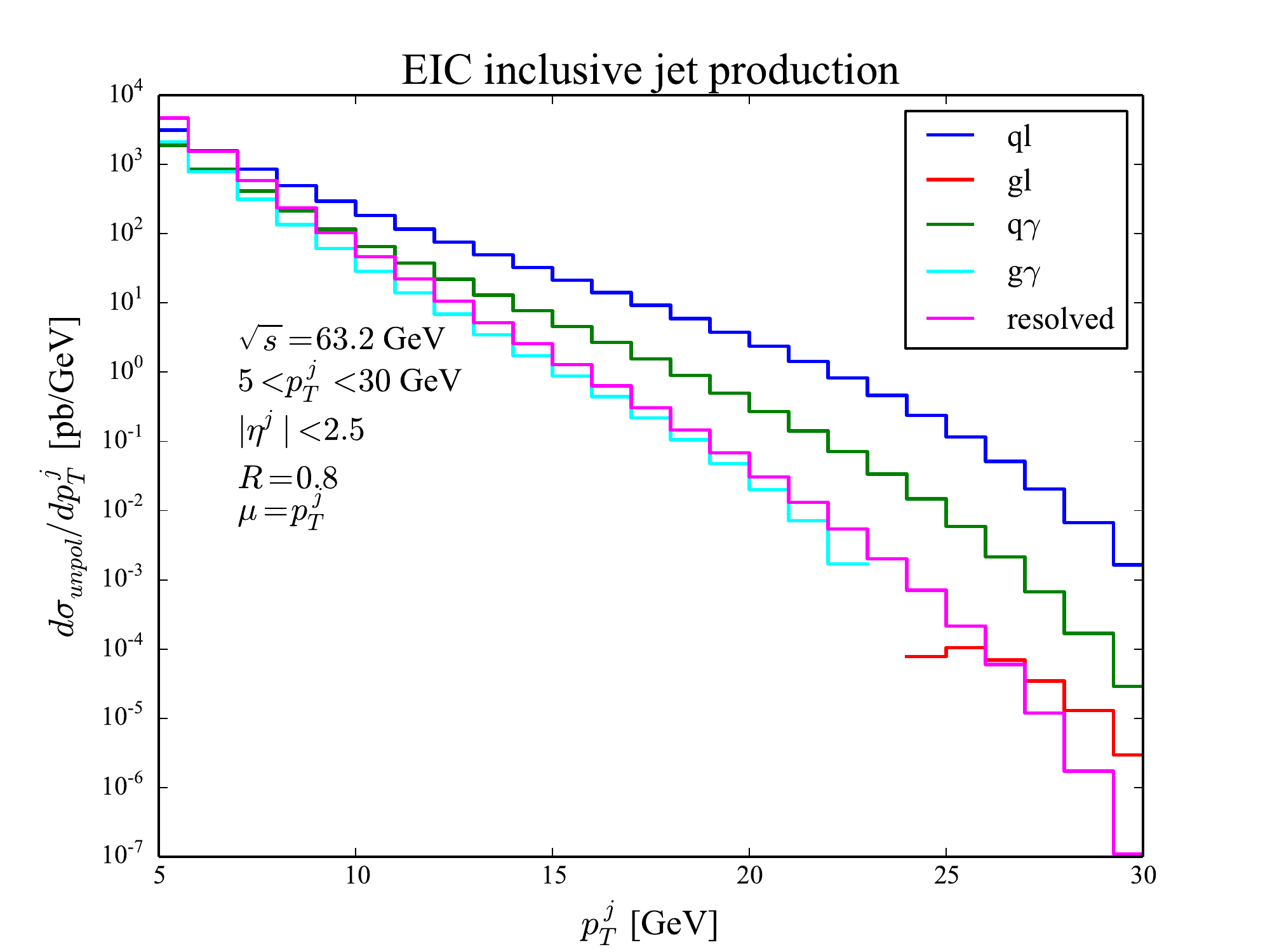, width=3.3in}
\hskip 0.2in
\psfig{file=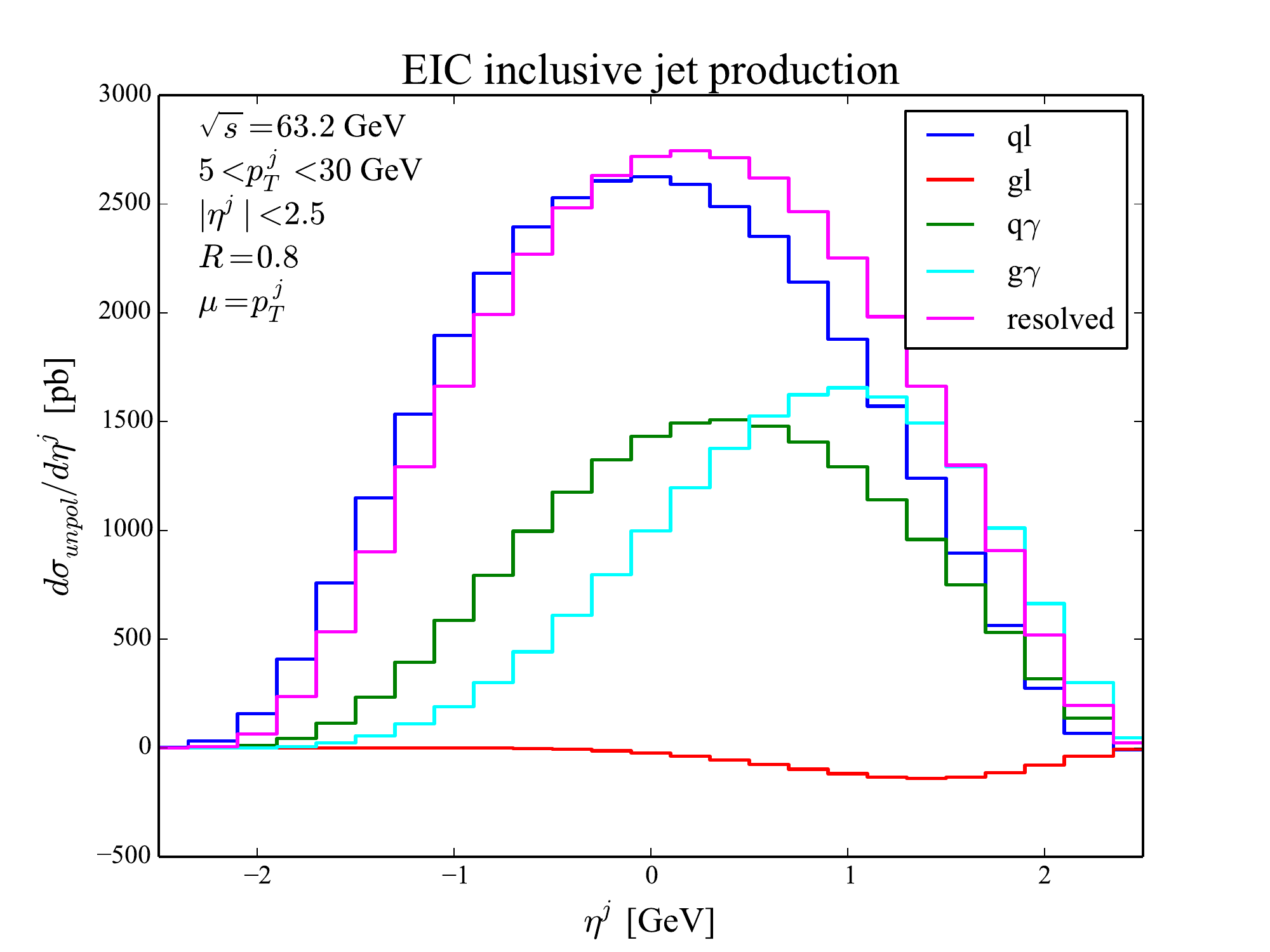, width=3.3in}
\caption{Split of the unpolarized transverse momentum and pseudorapidity distributions for $\sqrt{s}=63.2$ GeV into partonic channels as described in the text.}
\label{fig:sqrts63-totcr-split}
\eef

We now study the spin asymmetry for $\sqrt{s}=63.2$ GeV.  The distributions of $A_{LL}$ in both jet transverse momentum and pseudorapidity are shown in Fig.~\ref{fig:sqrts63-ALL}.  The separate partonic channels are shown in Fig.~\ref{fig:sqrts63-ALL-split}.  As before the separate contributions of both polarized resolved photon models are shown, as well as the result obtained by turning off the polarized gluon distribution.  The asymmetries are larger than observed for $\sqrt{s}=141.4$ GeV, reaching 40\% at large $p_T^j$ and nearly 3\% for central $\eta^j$ when integrated over transverse momentum.  The $ql$ channel dominates the asymmetry as a function of $p_T^j$ for all but the lowest few bins, where other partonic channels such as the resolved photon contributions, the $q\gamma$, and the $g\gamma$ terms all become important.  While the contribution of the $ql$ channel can be reduced by focusing on the low transverse momentum range $p_T^j \lesssim 10$ GeV, both the polarized gluon and resolved photon distributions become important for this region.  This is different than for $\sqrt{s}=141.4$ GeV, where the polarized gluon contributions could be more easily isolated.

The $\eta^j$ distribution again shows a difference in shape between the minimal and maximal models of the polarized photon distributions, although this channel is smaller than the $ql$ and $q\gamma$ ones throughout most of the phase space.  The forward $\eta^j$ region exhibits a strong interplay between the $q\gamma$ and $g\gamma$ channels.  They significantly cancel, indicating sensitivity to modifications of the polarized gluon distribution.  However, we note that the size of the $A_{LL}$ contribution coming from the $g\gamma$ channel is significantly smaller than observed for $\sqrt{s}=141.4$ GeV with $p_T^j>20$ GeV in Fig.~\ref{fig:sqrts141-ALL-pt20}, while the relative contribution from the resolved photon channel is larger, suggesting less sensitivity to this quantity than for the larger energy collisions. 

\begin{figure}[h]
\psfig{file=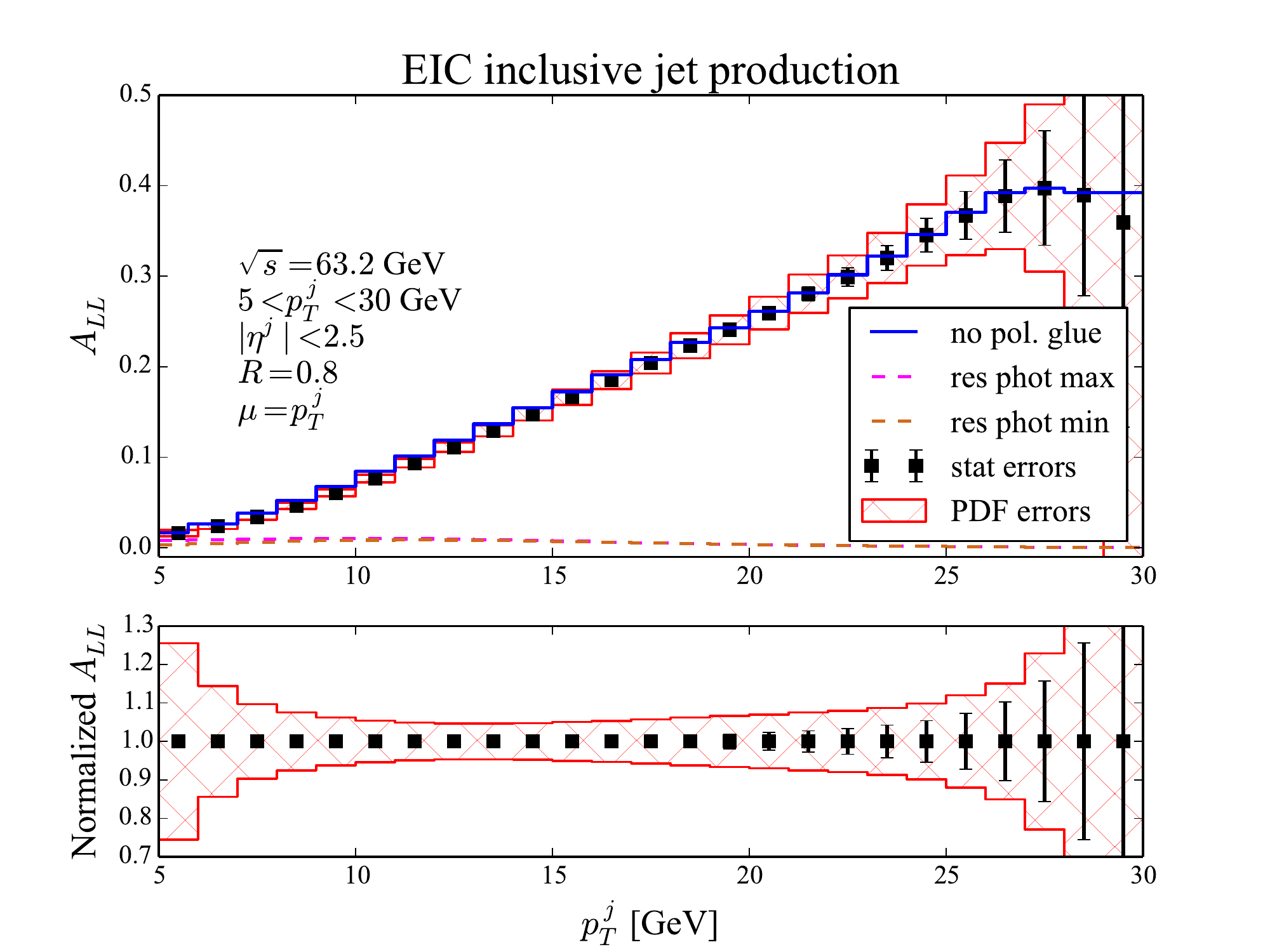, width=3.3in}
\hskip 0.2in
\psfig{file=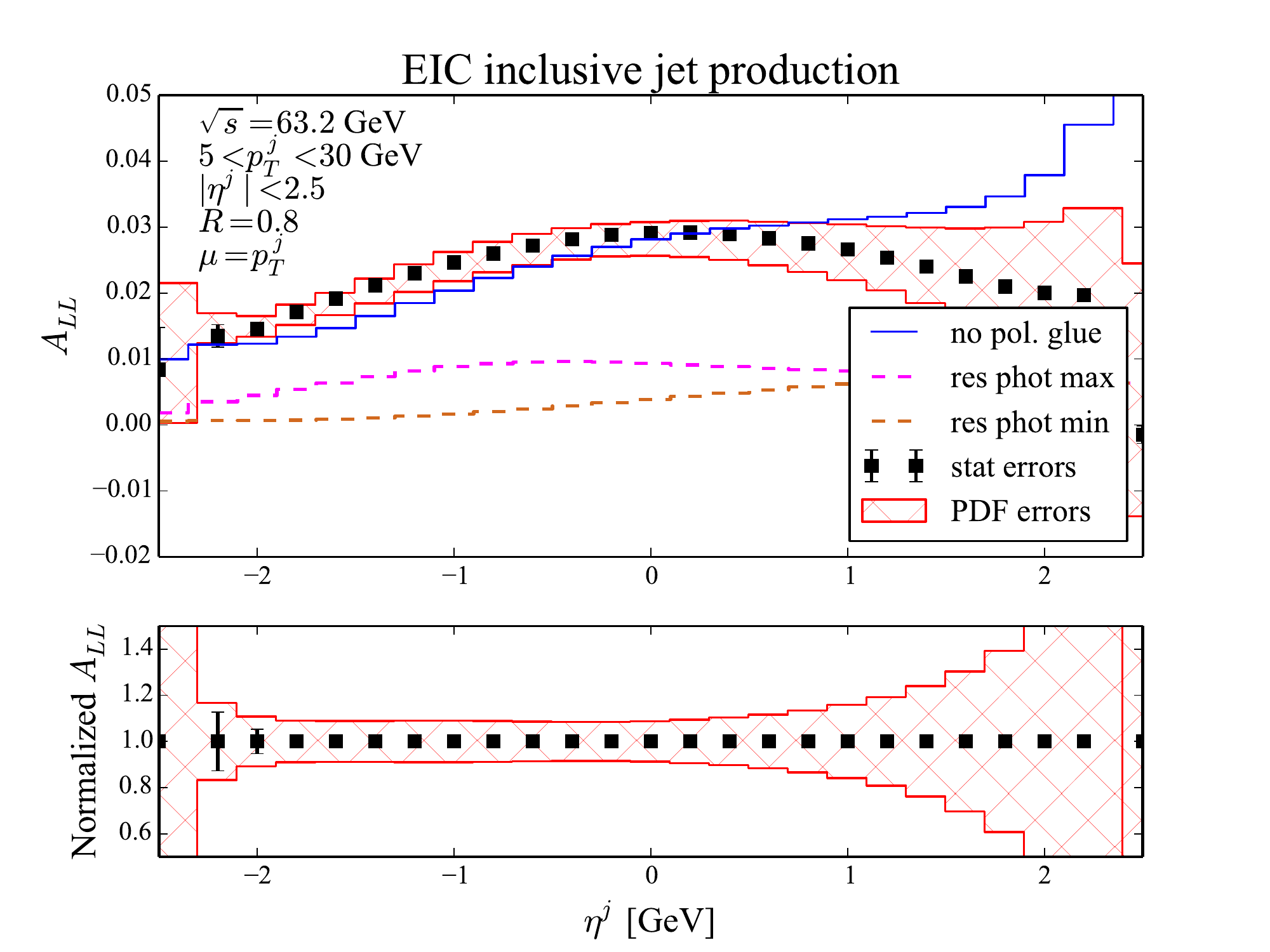, width=3.3in}
\caption{Spin asymmetry as a function of jet transverse momentum (left panel) and jet pseudorapidity (right panel) for $\sqrt{s}=63.2$ GeV. The resolved photon contribution is shown separately in the upper panel of each plot.  The lower panels normalize the results to the central values in order to more clearly illustrate the errors.}
\label{fig:sqrts63-ALL}
\eef

\begin{figure}[h]
\psfig{file=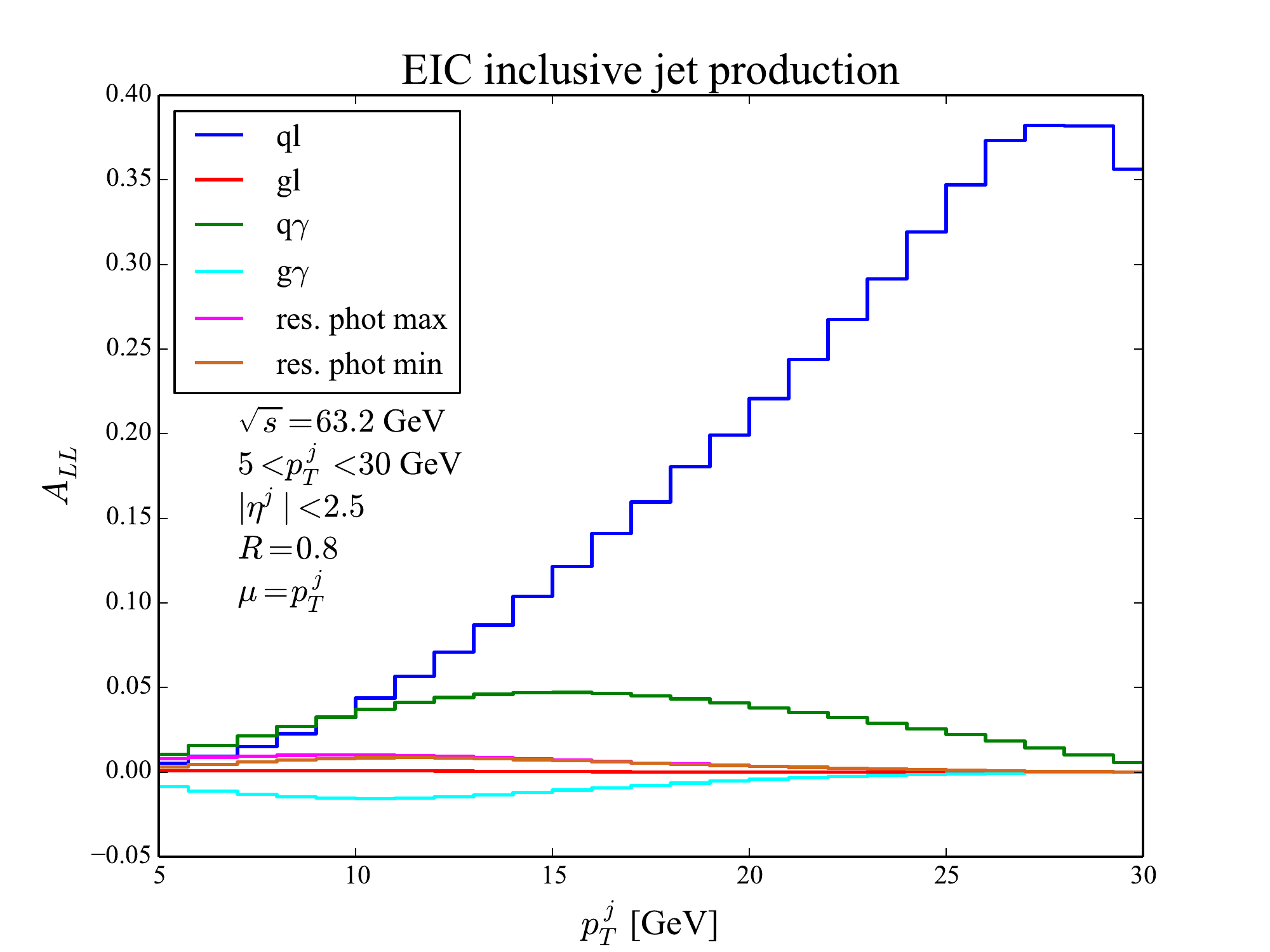, width=3.3in}
\hskip 0.2in
\psfig{file=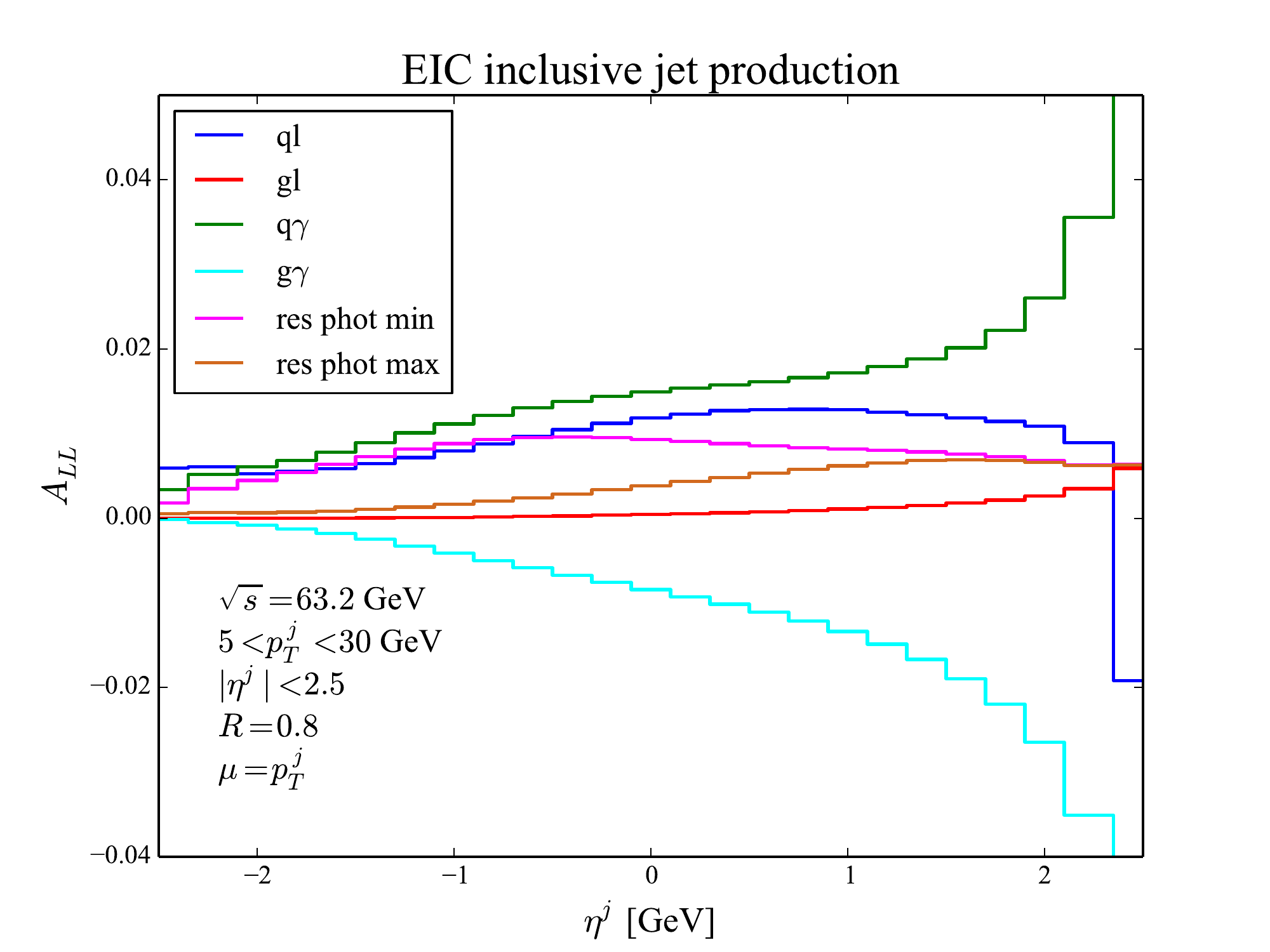, width=3.3in}
\caption{Splits of the spin asymmetry as functions of the jet transverse momentum and pseudorapidity  for $\sqrt{s}=63.2$ GeV into partonic channels.}
\label{fig:sqrts63-ALL-split}
\eef

\section{Results for $\sqrt{s} = 44.7$ GeV}
\label{sec:44}

Finally, we focus on our lowest considered center-of-mass scattering energy, $\sqrt{s}=44.7$ GeV.  We consider the kinematic range $5 < p_T^j< 20$ GeV and $|\eta^j|<2$.  We first show in Fig.~\ref{fig:sqrts44-totcr} the total cross section as a function of transverse momentum and pseudorapidity.  The splits of these distributions into separate partonic channels are shown in Fig.~\ref{fig:sqrts44-totcr-split}.  The most prominent difference with respect to the previously studied energies is that the resolved photon contributions are no longer dominant in any region of phase space.  The $ql$ channel is the largest over the entire studied region, followed by the $q\gamma$ channel.  From the perspective of better determining the partonic structure of the photon via jet production, it is advantageous to be at higher collisions energies.

\begin{figure}[h]
\psfig{file=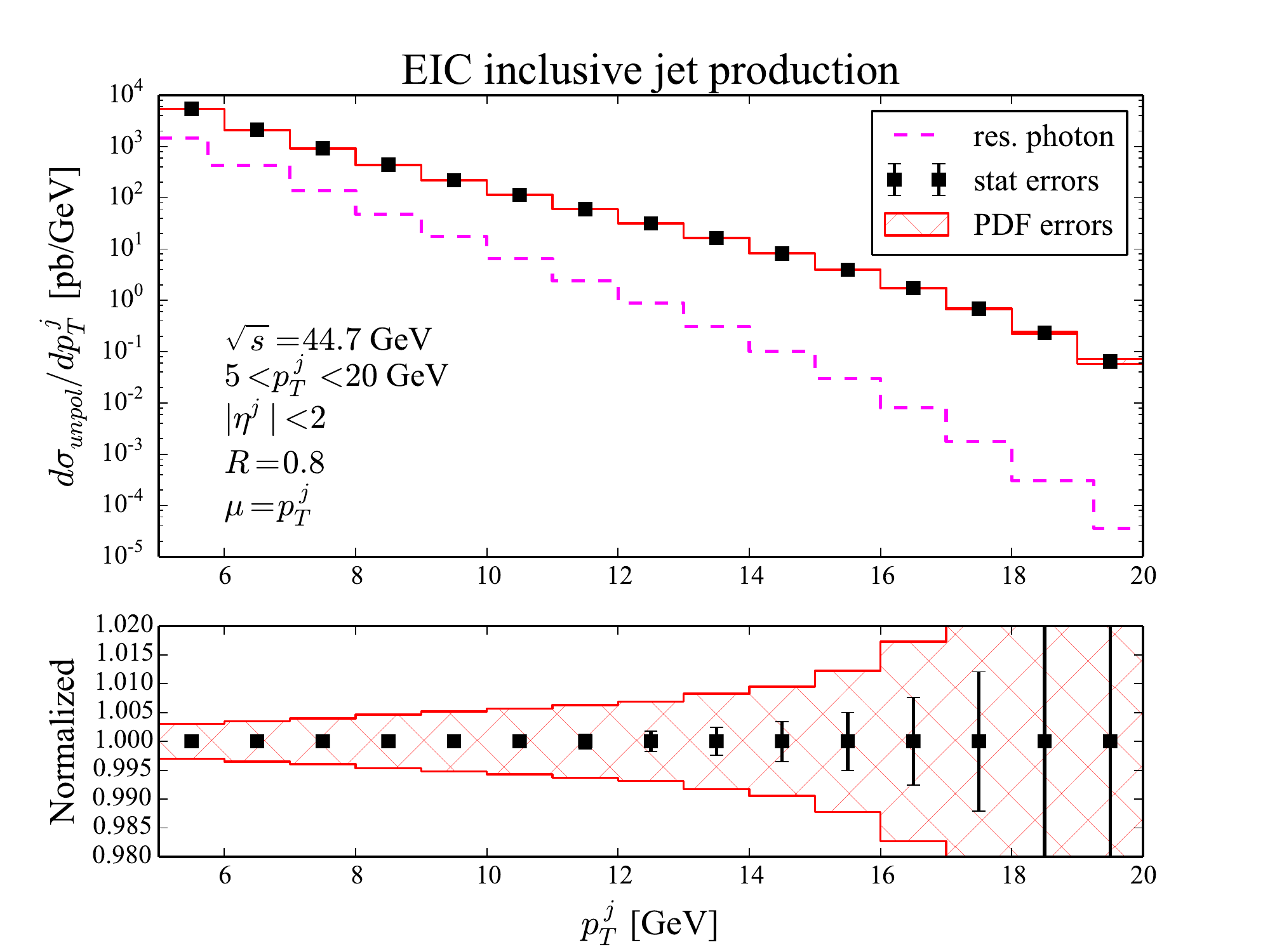, width=3.3in}
\hskip 0.2in
\psfig{file=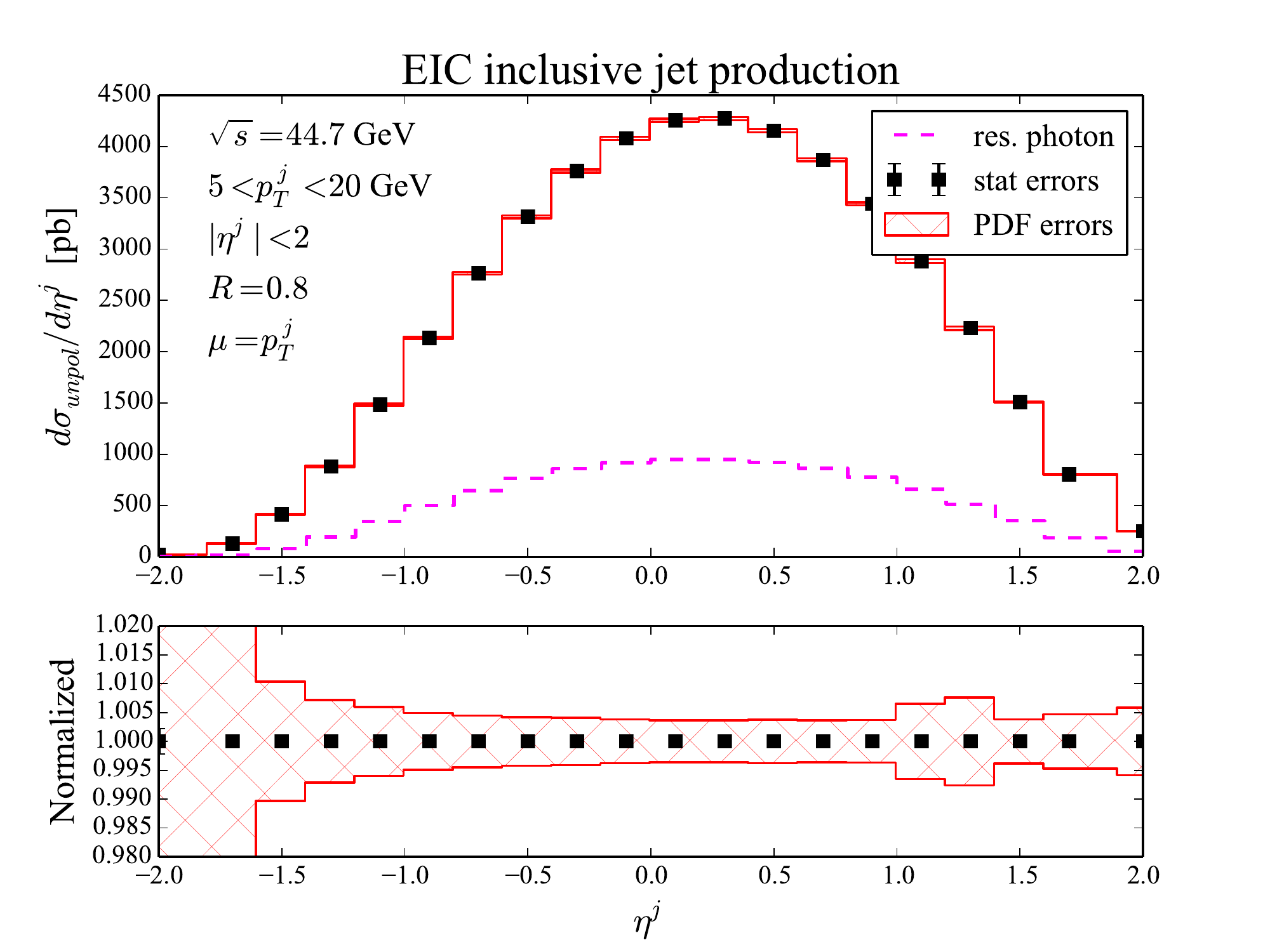, width=3.3in}
\caption{Total unpolarized cross section as a function of jet transverse momentum (left panel) and jet pseudorapidity (right panel) for $\sqrt{s}=44.7$ GeV. The resolved photon contribution is shown separately in the upper panel of each plot.  The lower panels normalize the results to the central values in order to more clearly illustrate the errors.}
\label{fig:sqrts44-totcr}
\eef

\begin{figure}[h]
\psfig{file=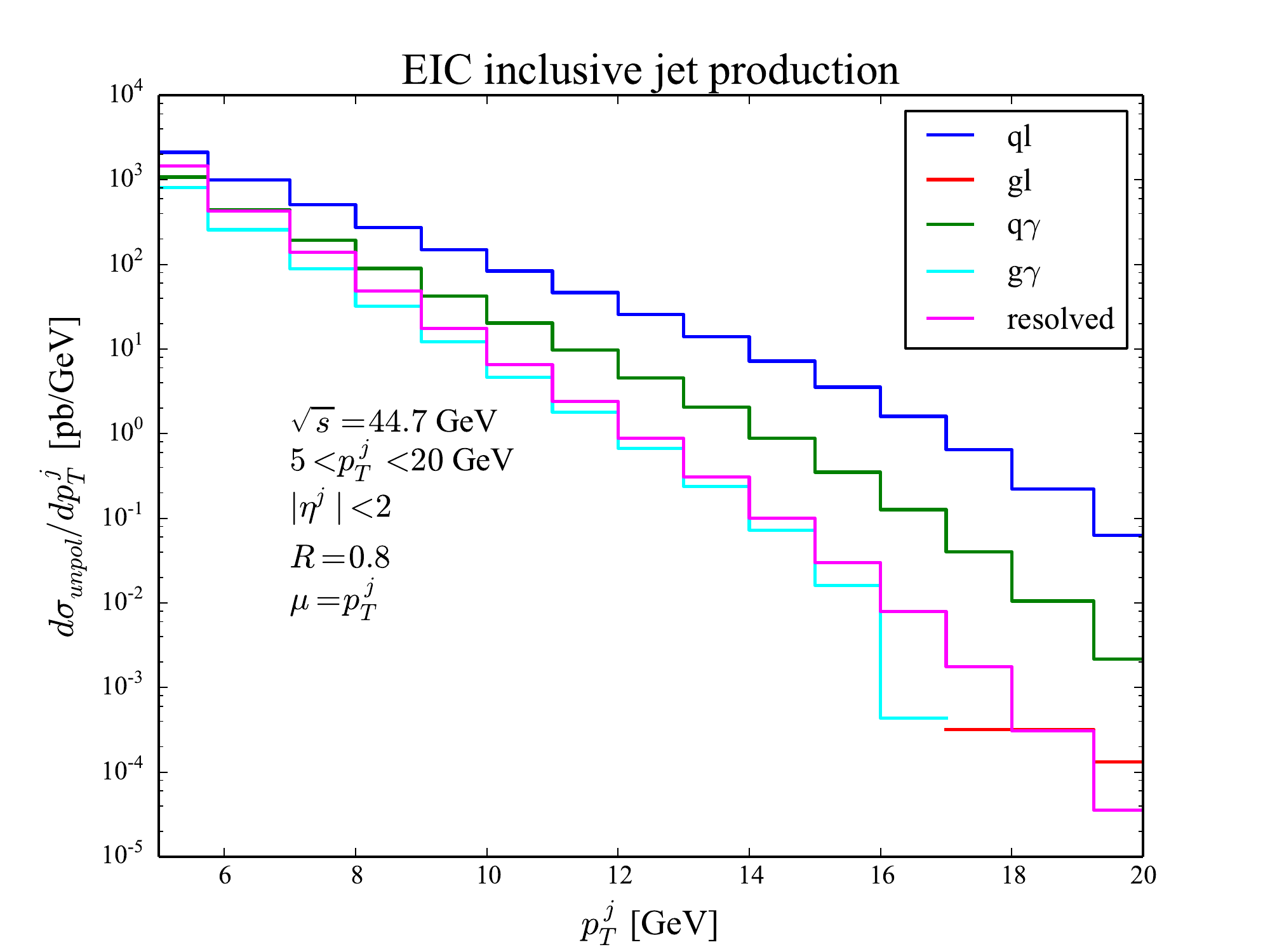, width=3.3in}
\hskip 0.2in
\psfig{file=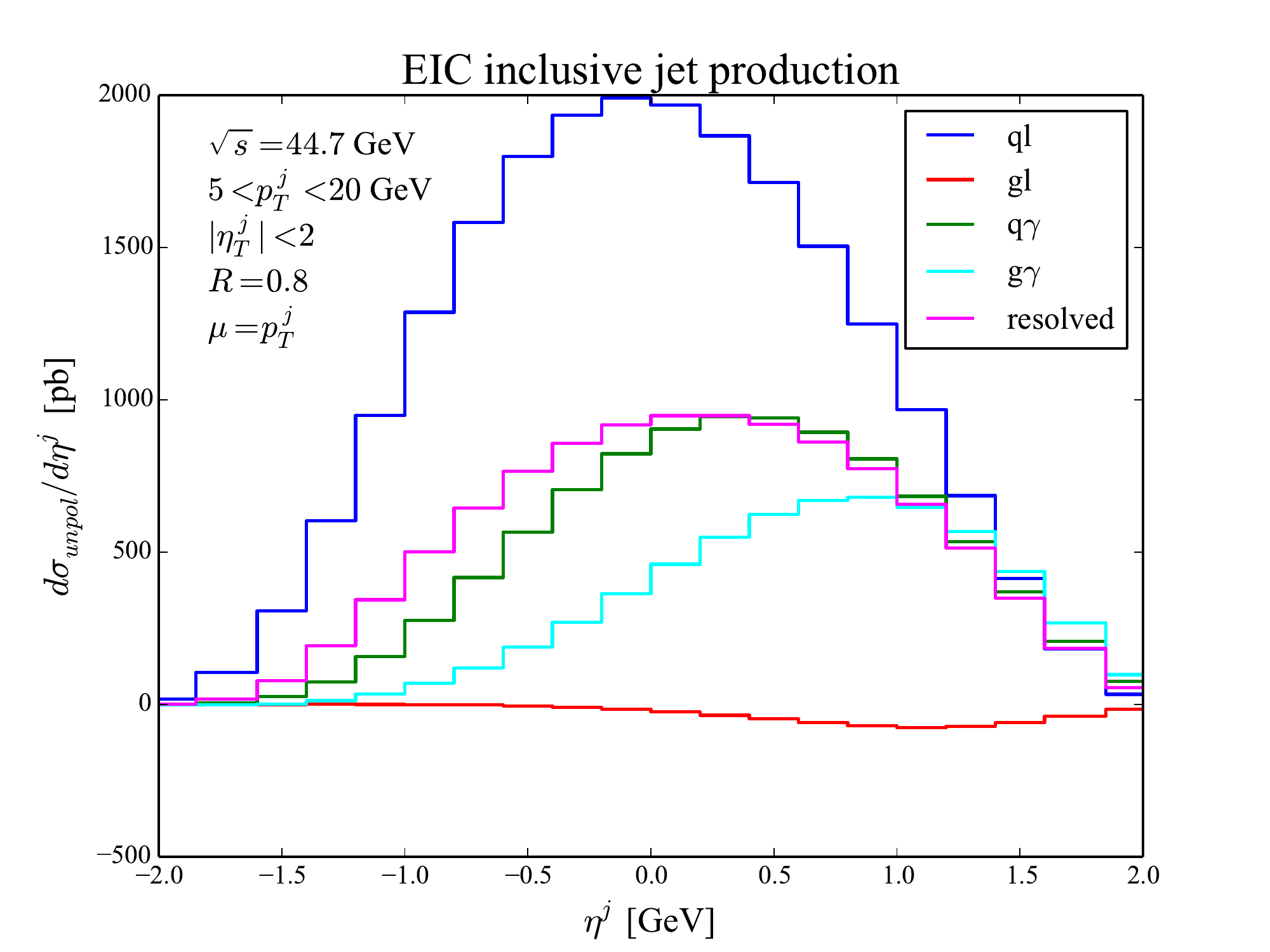, width=3.3in}
\caption{Split of the unpolarized transverse momentum and pseudorapidity distributions for $\sqrt{s}=44.7$ GeV into partonic channels as described in the text.}
\label{fig:sqrts44-totcr-split}
\eef

In Fig.~\ref{fig:sqrts44-ALL} we show the spin asymmetry $A_{LL}$ as a function of $p_T^j$ and $\eta^j$.  The contributions of the separate partonic channels are shown in Fig.~\ref{fig:sqrts44-ALL-split}.  We first note that the estimated PDF errors are much smaller than for larger center-of-mass energies, reaching only a maximum of 20\% at the boundaries of phase space.  This is in contrast to $\sqrt{s}=141.4$ GeV, where Fig.~\ref{fig:sqrts141-ALL} exhibits PDF errors reaching 40\% or more for $\eta^j>0$ and $p_T^j<10$ GeV.  As Fig.~\ref{fig:xrange} makes clear, collisions at $\sqrt{s}=44.7$ GeV probe large Bjorken-$x$ where some knowledge of the polarized structure of the proton is available.  Higher collision energies probe smaller-$x$, which are still undetermined from data.  Broader coverage of the polarized PDFs is obtained by measurements at higher scattering energy.  While there are differences between the various models for the polarized resolved photons, both models give significantly smaller contributions than the $ql$ channel throughout phase space.  There is a significant cancellation between the $q\gamma$ and $g\gamma$ contributions to the asymmetry in the forward $\eta^j$ region, but these separate contributions to the asymmetry only reach a few percent, smaller than observed for $\sqrt{s}=141.4$ GeV with the cut $p_T^j>20$ GeV.

\begin{figure}[h]
\psfig{file=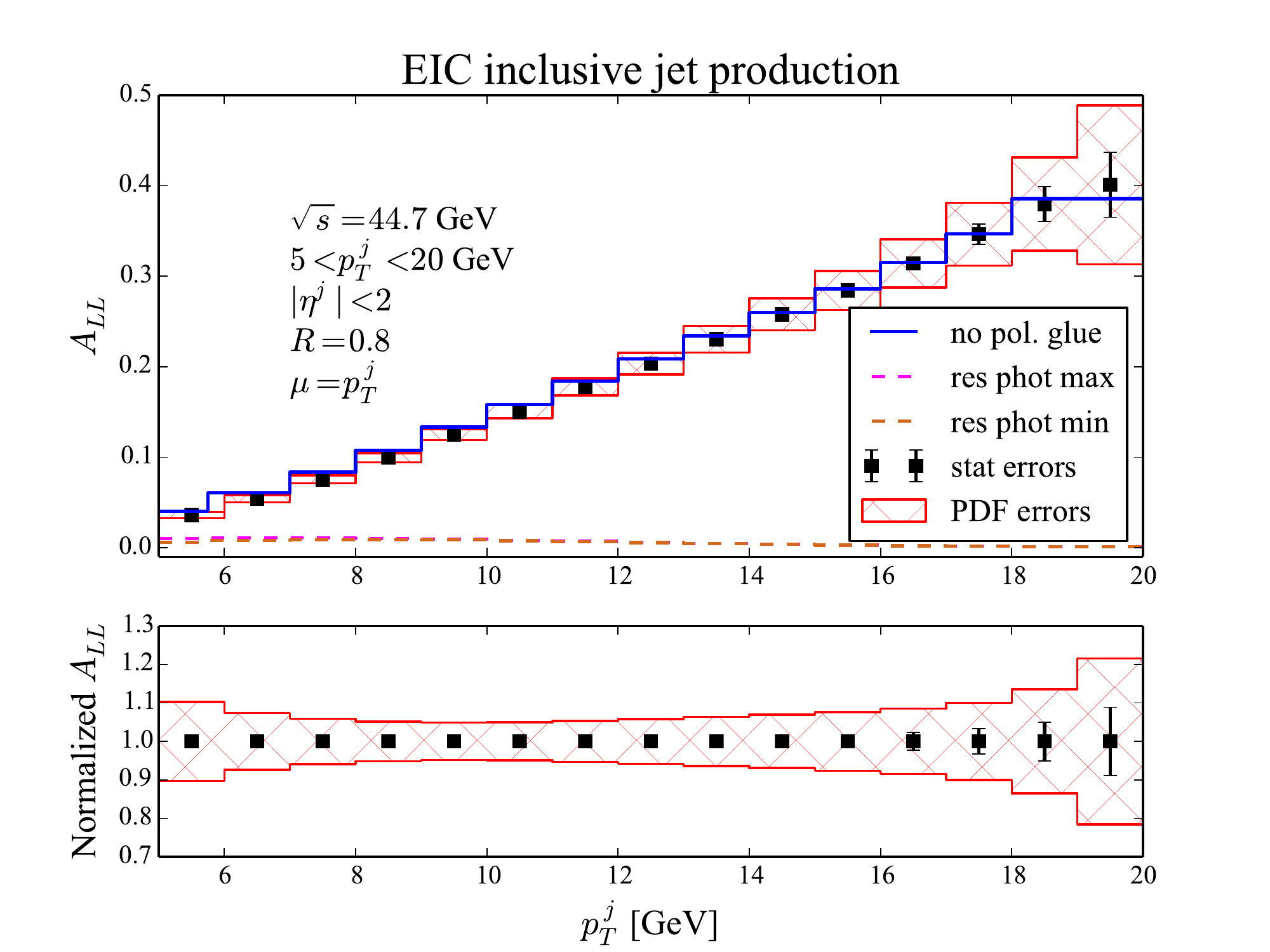, width=3.3in}
\hskip 0.2in
\psfig{file=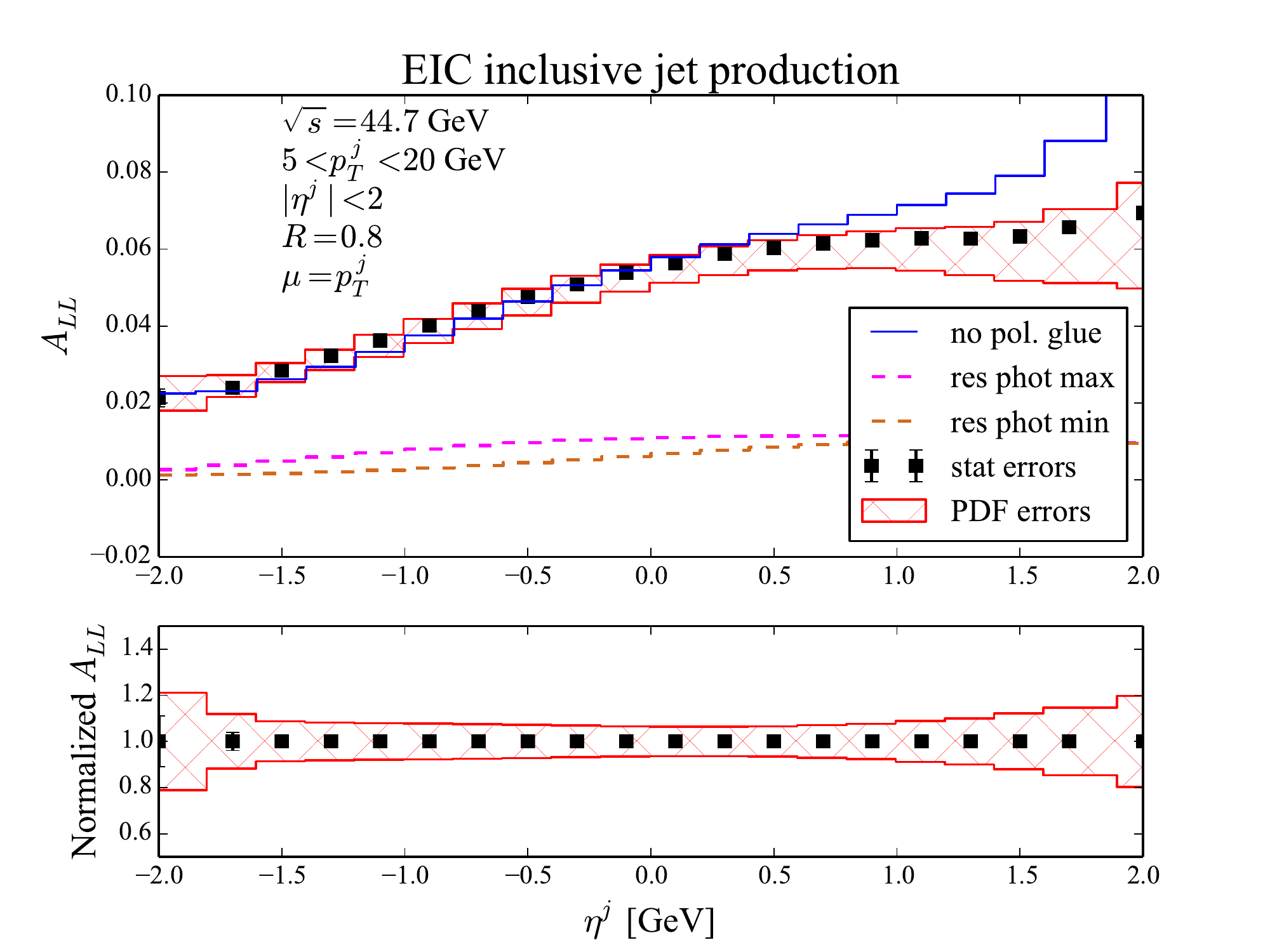, width=3.3in}
\caption{Spin asymmetry as a function of jet transverse momentum (left panel) and jet pseudorapidity (right panel) for $\sqrt{s}=44.7$ GeV. The resolved photon contribution is shown separately in the upper panel of each plot.  The lower panels normalize the results to the central values in order to more clearly illustrate the errors.}
\label{fig:sqrts44-ALL}
\eef

\begin{figure}[h]
\psfig{file=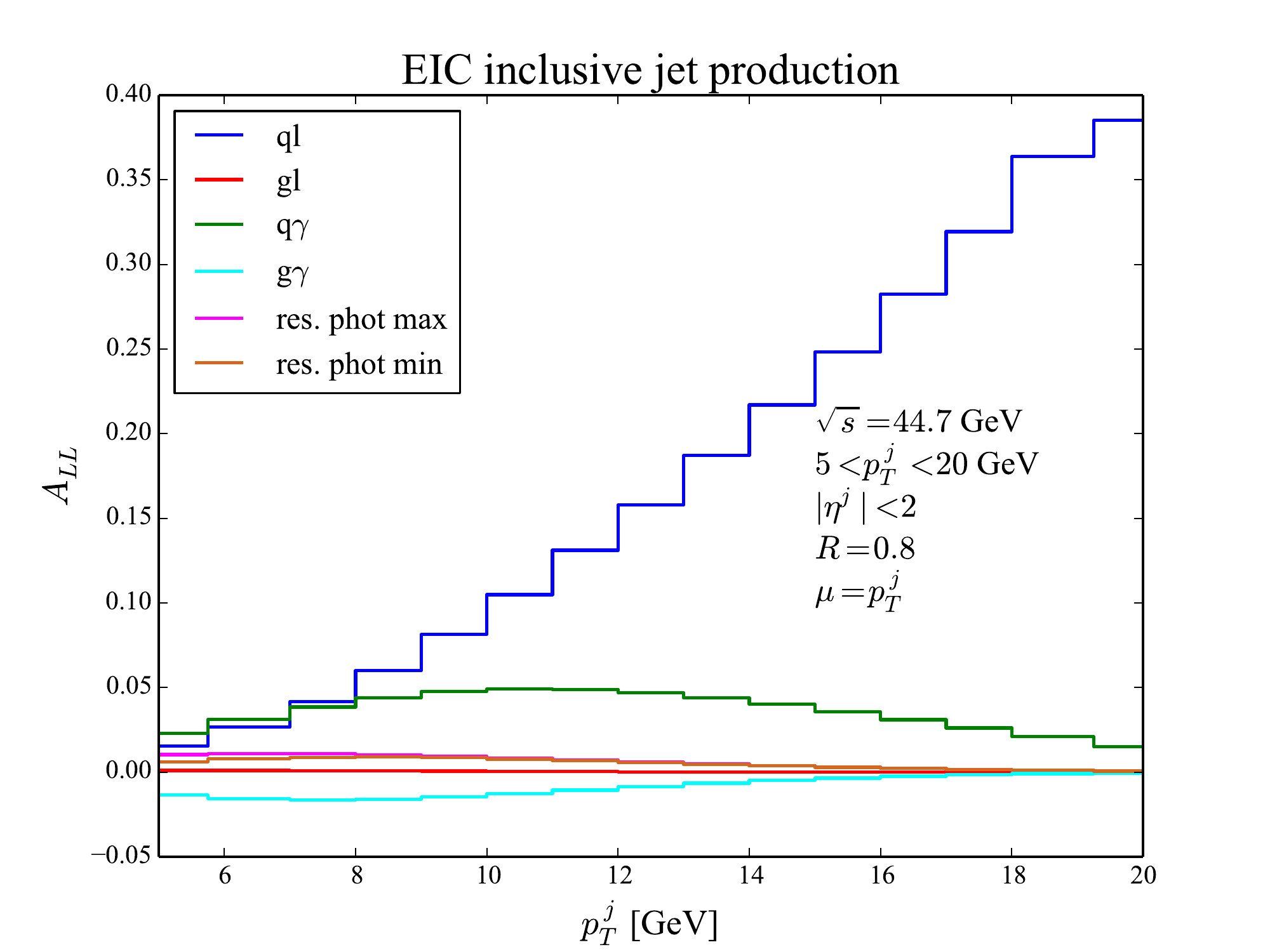, width=3.3in}
\hskip 0.2in
\psfig{file=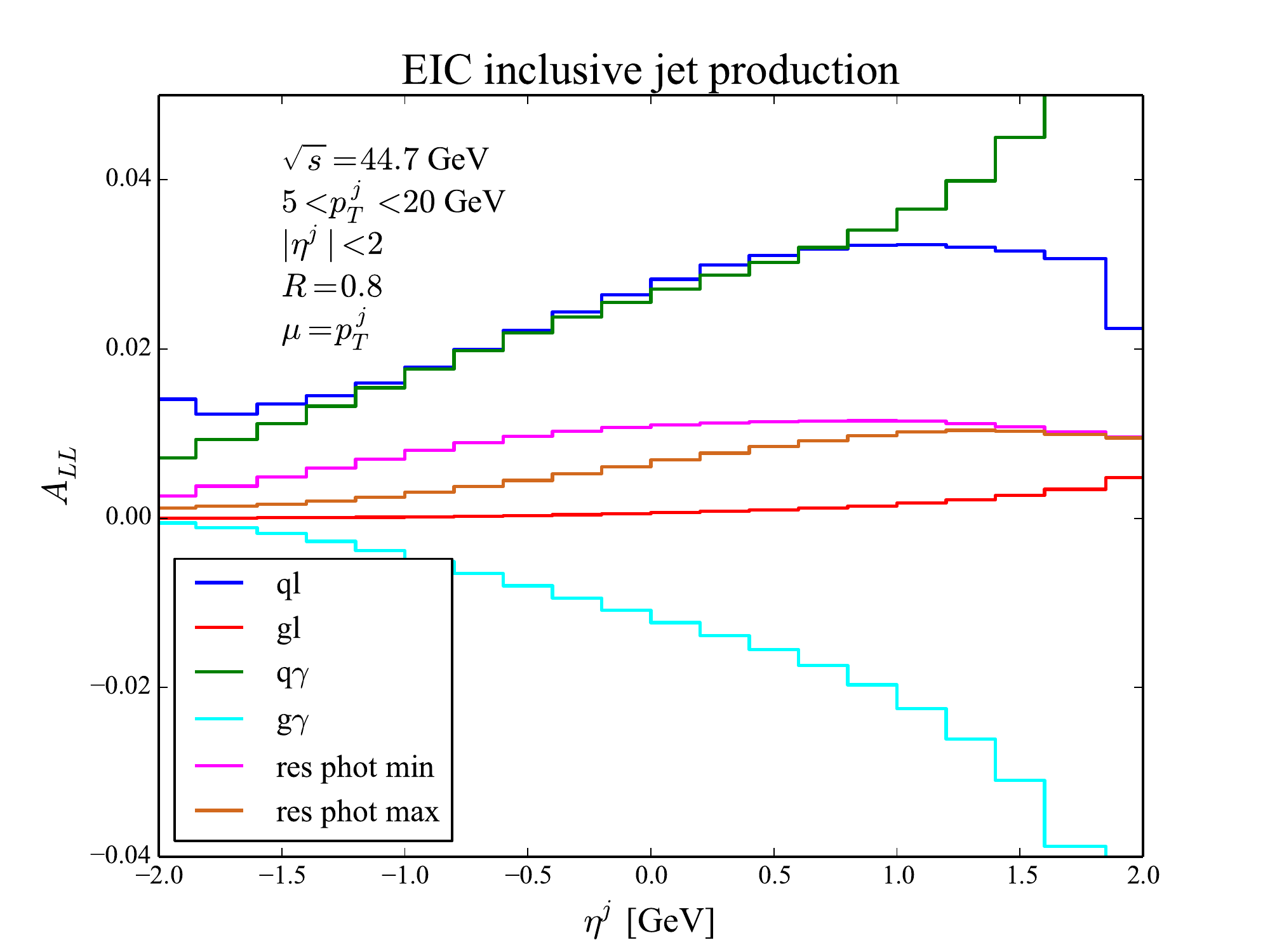, width=3.3in}
\caption{Splits of the spin asymmetry as functions of the jet transverse momentum and pseudorapidity  for $\sqrt{s}=44.7$ GeV into partonic channels.}
\label{fig:sqrts44-ALL-split}
\eef

\section{Summary and Conclusions}
\label{sec:conc}

In this manuscript we presented a detailed phenomenological study of inclusive jet production at a future EIC.  Our goal was to establish the ability of this process to probe the polarized structure of the proton, and to determine which kinematic regions of jet production are sensitive to different aspects of proton structure.  We have considered several possibilities for the center-of-mass scattering energy of the proposed machine to elucidate how different EIC realizations can improve our knowledge of proton structure.  We have studied the effects of different PDF parameterizations, finite jet radii and the effect of tagging the final-state lepton.  Both the polarized proton and photon structures were considered in our analysis.  The effects of statistical and PDF uncertainties were studied in detail.  

Our study was performed using fixed-order perturbative QCD through NLO with all relevant partonic processes included, including the resolved contributions associated with the non-perturbative structure of the photon.  The entire theoretical framework for both polarized and unpolarized collisions has been implemented in the flexible numerical code {\tt DISTRESS} designed for phenomenological studies at a future EIC.  The major findings of our study are summarized below.
\begin{itemize}

\item Collisions at the highest considered center-of-mass energy, $\sqrt{s}=141.4$ GeV, offer the broadest sensitivity to polarized hadronic structure.  Both the resolved photon distributions and the polarized gluons and quarks can be probed by selecting appropriate regions of jet transverse momentum and pseudorapidity.  Low transverse momenta provide access to the resolved photon, while intermediate-to-high transverse momenta are sensitive to the polarized gluon.  As the scattering energy is decreased the sensitivity to the polarized photon distributions decreases, since these distributions fall rapidly as higher Bjorken-$x$ values are probed in lower-energy collisions.  In particular at the lowest studied energy, $\sqrt{s}=44.7$ GeV, it is difficult to access these distributions. 

\item The estimated polarized PDF errors are much larger than the expected EIC statistical errors.  Previous work~\cite{Abelof:2016pby} has shown that the theoretcial scale uncertainties are small once NNLO corrections are included.  There is an excellent chance to learn more about polarized proton structure at the EIC. 

\item The sensitivity to the polarized gluon comes from scattering channels such as $g\gamma$ which become active when $Q^2 \approx 0$.  Large polarized quark contributions with significant PDF errors comes from the $ql$ scattering channel at high transverse momentum.  Resolved photon distributions can be accessed at low transverse momentum.  This shows the importance of inclusive jet production, since it gives handles on all relevant distributions in different kinematic regions.

\end{itemize}

\section*{Acknowledgments}
We thank W.~Vogelsang for comments on the manuscript.  R.~B. is supported by the DOE contract DE-AC02-06CH11357.  F.~P. is
supported by the DOE grants DE-FG02-91ER40684 and DE-AC02-06CH11357.
H.~X. is supported by the DOE grant DE-AC02-06CH11357 and the NSF
grant NSF-1740142.  This research used resources of the Argonne
Leadership Computing Facility, which is a DOE Office of Science User
Facility supported under Contract DE-AC02-06CH11357.


\end{document}